\documentclass[useAMS,usenatbib]{mn2e}
\usepackage{xspace}
\usepackage{amssymb}
\usepackage{graphicx} 
\usepackage[T1]{fontenc}
\newcommand{\sz}{$\sigma_z$\xspace}
\newcommand{\sr}{$\sigma_r$\xspace}

\newcommand{\kms}{km~s$^{-1}$\xspace} 
\newcommand{\msun}{M$_{\odot}$\xspace}

\title{Dissecting simulated disc galaxies II: the age-velocity relation}

\author[M. Martig, I. Minchev and C. Flynn]{ Marie Martig$^{1,2}$\thanks{E-mail: marie.martig@gmail.com}, Ivan Minchev$^{3}$ and Chris Flynn$^{1}$\\
$^1$Centre for Astrophysics \& Supercomputing, Swinburne University of Technology, P.O. Box 218, Hawthorn, VIC 3122, Australia \\
$^2$Max-Planck-Institut f\"{u}r Astronomie, K\"{o}nigstuhl 17, 69177 Heidelberg, Germany\\
$^{3}$Leibniz-Institut f\"{u}r Astrophysik Potsdam (AIP), An der Sternwarte 16, 14482, Potsdam, Germany
}
\begin{document}
\date{Accepted 2014 June 23.  Received 2014 June 20; in original form 2013 September 12}
\maketitle
\begin{abstract} 
We study the relation between stellar ages and vertical velocity dispersion (the age-velocity relation, or AVR) in a sample of seven simulated disc galaxies. In our simulations, the shape of the AVR for stars younger than 9 Gyr depends strongly on the merger history at low redshift, with even 1:10 -- 1:15 mergers being able to create jumps in the AVR (although these jumps might not be detectable if the errors on stellar ages are on the order of 30\%). For galaxies with a quiescent history at low redshift, we find that the vertical velocity dispersion rises smoothly for ages up to 8--9 Gyr, following a power law with a  slope of $\sim 0.5$, similar to what is observed in the solar neighbourhood by the Geneva-Copenhagen Survey. For these galaxies, we show that the slope of the AVR is not imprinted at birth, but is the result of subsequent heating. By contrast, in all our simulations, the oldest stars form a significantly different population, with a high velocity dispersion. These stars are usually born kinematically hot in a turbulent phase of intense mergers at high redshift, and also include some stars accreted from satellites. This maximum in \sz is strongly decreased when age errors are included, suggesting that observations can easily miss such a jump with the current accuracy of age measurements.
\end{abstract}


\section{Introduction} 

In the solar neighbourhood, the vertical velocity dispersion (\sz) of stars is correlated with their age. Stars with an age of 9--10 Gyr have a \sz of about 40 \kms (or more, depending on the selection criteria, and in particular on the inclusion or exclusion of halo stars, see for instance \citealp{Casagrande2011}). By contrast, 1 Gyr-old stars have a \sz of 12--16 \kms, and the youngest stars are rather cold: \cite{Aumer2009} find that the bluest stars have a \sz of only 6 \kms.

\begin{figure}
\centering 
\includegraphics[width=0.45\textwidth]{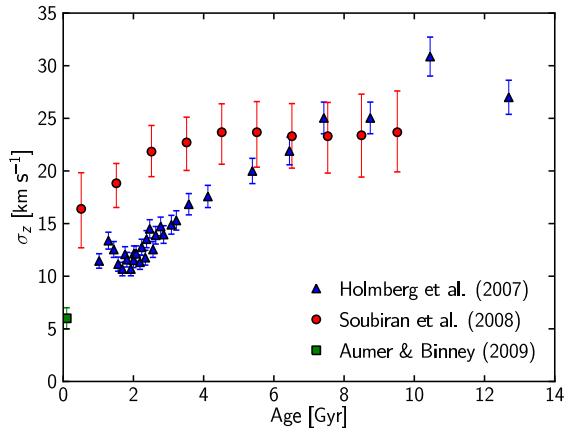}
\caption{The age-velocity relation observed in the solar neighbourhood: blue triangles from Holmberg et al. (2007), red dots from Soubiran et al. (2008). The green square corresponds to the bluest stars from Aumer \& Binney (2009), arbitrarily placed at a very small age. This illustrates the two competing ways of seeing the shape of the AVR: with a saturation after a few Gyrs (as measured by Soubiran et al.), or with a smooth increase of \sz with age (as measured by Holmberg et al.).}
\label{fig:observations}
\end{figure}

The exact shape of the age-velocity relation (or AVR) is however debated, partly because measuring stellar ages is difficult \citep{Soderblom2010}, and partly because the selection criteria differ between different surveys \citep{Haywood2013}: we show in Figure \ref{fig:observations} two examples of observational results giving AVRs with very different shapes.
For instance, \cite{Quillen2001} propose a rapid increase of \sz for ages smaller than 3 Gyr, followed by a plateau corresponding to a saturation of \sz at $\sim$ 20 \kms, and a final jump for ages greater than 9 Gyr. \cite{Soubiran2008} find a similar saturation, but occuring for stars older than 5 Gyr. By contrast, others argue in favour of a continuously rising \sz with age, following a power law \sz$\propto t^{\alpha}$ where $\alpha$ is close to 0.5 \citep[e.g.,][]{Wielen1977,Nordstrom2004,Holmberg2007, Aumer2009}. However, \cite{Seabroke2007} as well as \cite{Aumer2009} have shown that the current data do not favour a power law evolution over a saturation of \sz for ages greater than 4--5 Gyr. The saturation after 3 Gyr that was proposed by \cite{Quillen2001} from a sample of 189 stars is excluded by the current data.

A possibility is that the shape of the AVR is determined by the properties of stars at birth, i.e. stars are born with a decreasing \sz with decreasing redshift. This is supported by observations of turbulent gas discs at high redshift \citep[e.g.,][]{Forster2009}, so that stars are naturally born hotter at higher redshift (note however that typical clumpy galaxies might be too massive to be the progenitors of Milky Way-type galaxies, e.g. \citealp{vanDokkum2013,Inoue2014}), and by the idea that turbulence (and thus \sz) decreases with redshift because of decreasing gas accretion rates \citep{Bournaud2009,Forbes2012,Bird2013}.

Most theoretical works have assumed that stars were born with a roughly constant \sz over the last 8--9 Gyrs, and that their present-day \sz is the result of gradual heating. \cite{Spitzer1951,Spitzer1953} first proposed that the source of that heating could be the scattering of stars by giant molecular clouds with masses of $10^6$ \msun. It was later shown that the number of GMCs in the Milky Way is too small to reproduce the observed heating, and that ratio of vertical to radial velocity dispersions is too high compared to observations \citep{Lacey1984}.

By contrast, spiral arms are very inefficient at vertical heating but can provide some radial heating provided that they are transient and stochastic \citep{Carlberg1985}, or that multiple density waves coexist  \citep{Minchev2006}. It was thus suggested that heating by a combination of spiral arms and GMCs could match the observed AVR \citep{Carlberg1987, Jenkins1990}. However, \cite{Hanninen2002} showed that heating by GMCs is not as efficient as previously thought, and instead proposed a combination of heating by GMCs and halo black holes (see also \citealp{Lacey1985} for an early study of heating by halo black holes).

Another important mechanism for disc evolution is the radial migration of stars under the action of non-axisymmetric perturbations \citep[e.g.,][]{Sellwood2002, Minchev2006,Minchev2010}. In recent years the idea has emerged that radial migration could be a source of disc heating: the analytical model of \cite{Schonrich2009a,Schonrich2009b} combines recipes for radial migration and chemical evolution and shows that migration alone could be responsible for the chemo-dynamical trends observed in the Milky Way. As a result, disc heating has been attributed to radial migration in the simulations analysed by \cite{Loebman2011}. However, the model by  \cite{Schonrich2009a,Schonrich2009b} assumes that the vertical energy is conserved during migration, while simulations have later shown that it is the vertical action that is conserved \citep{Solway2012,Minchev2012}. The important consequence is that the contribution of radial migration to disc heating is negligible, except in the outer regions where it induces some flaring \citep{Minchev2012}. This result is the consequence of the balance between outwards migrating stars, which heat the disc, and inward migrators, which  cool it (note that \citealp{Roskar2013} find that migration induces heating but only consider outwards migrators). 

It thus seems that while internal secular evolution clearly induces disc thickening, it is often difficult to figure out which mechanisms are acting in realistic simulations that include gas, spiral arms, bars and live dark matter halos (see for instance \citealt{Saha2010}, for simulations where heating in the inner disc is due to a bar, but where the causes of heating in the outer disc are less clear). In any case, it would seem that radial migration is not a dominant contributor to secular heating.

For real galaxies however a final potentially important source of heating is the interaction with satellite galaxies, mostly in the form of minor mergers \citep[e.g.,][]{Quinn1993,Walker1996, Villalobos2008, Kazantzidis2009}. The amount of heating provided depends not only on the merger ratio, but also on many other parameters like the orbit of the satellite \cite{Velazquez1999} or the gas content of the main galaxy \citep{Moster2010}.

\cite{House2011} have shown from a sample of cosmological simulations performed with different techniques that mergers create jumps in the AVR, and thus point towards a quiescent merger history for the Milky Way, for which no such jumps are seen for ages lower than 9 Gyr.

This paper is the second of a series that uses a sample of seven simulated galaxies with different merger histories to understand the effect of mergers on the structure of galactic discs. In Paper I \citep{Martig2014a}, we present in detail our sample and the analyses we perform.  We dissect the discs in so-called mono-age populations (in 500 Myr-wide age bins), and compare their structure to Milky Way observations by \cite{Bovy2012a,Bovy2012b,Bovy2012c}. In this paper, we focus solely on the AVR. 
We present briefly our simulations in Section 2, and then discuss the velocity dispersion of young stars in Section 3. In Section 4, we compare the AVR in galaxies with quiescent or active merger histories. In Section 5, we study the origin of the slope of the AVR in quiescent galaxies, and discuss potential numerical limitations in Section 6 before summarizing our results in Section 7.

\section{Simulations and analysis}

\begin{figure*}
\centering 
\includegraphics[width=0.245\textwidth]{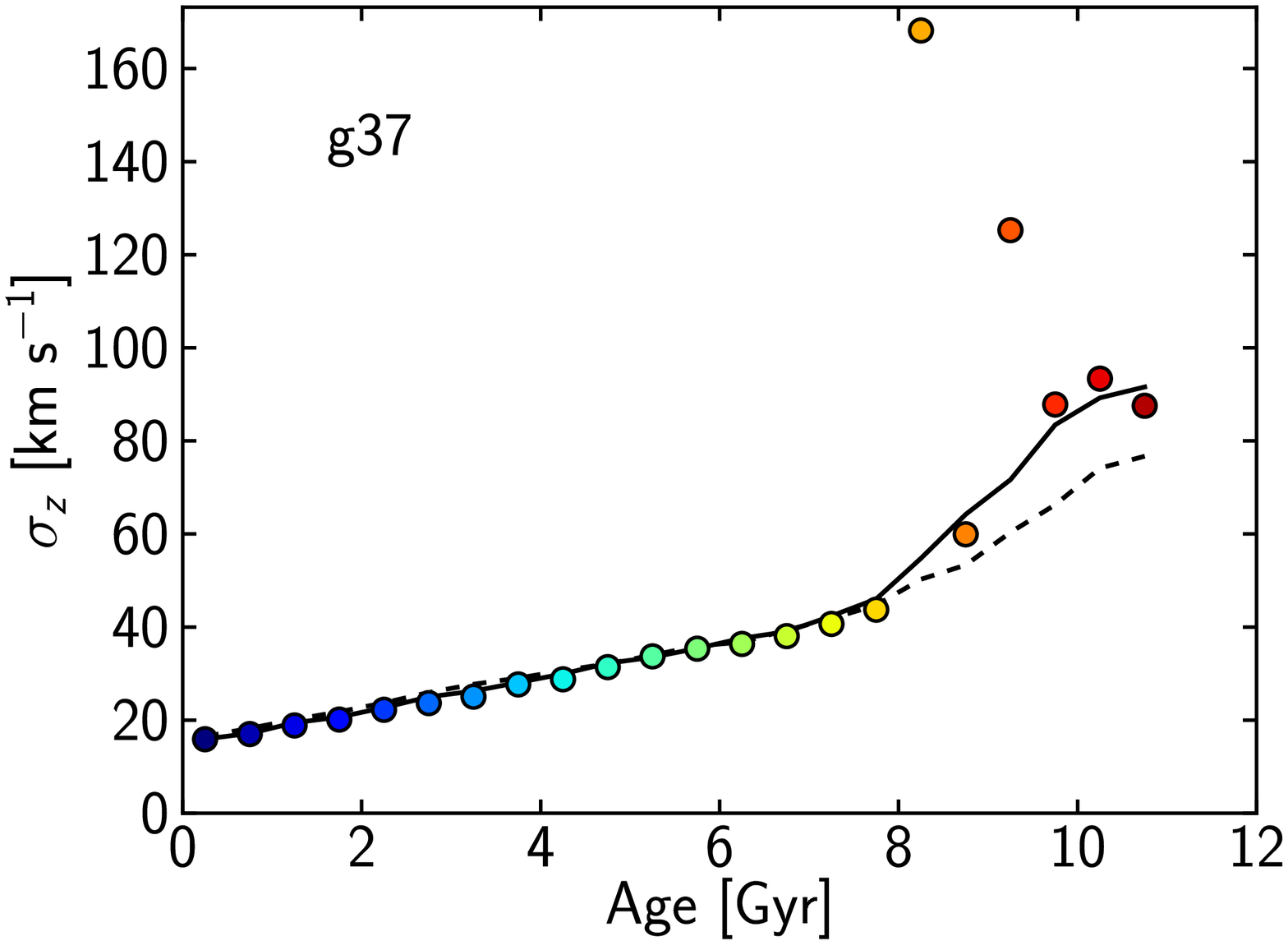}
\includegraphics[width=0.245\textwidth]{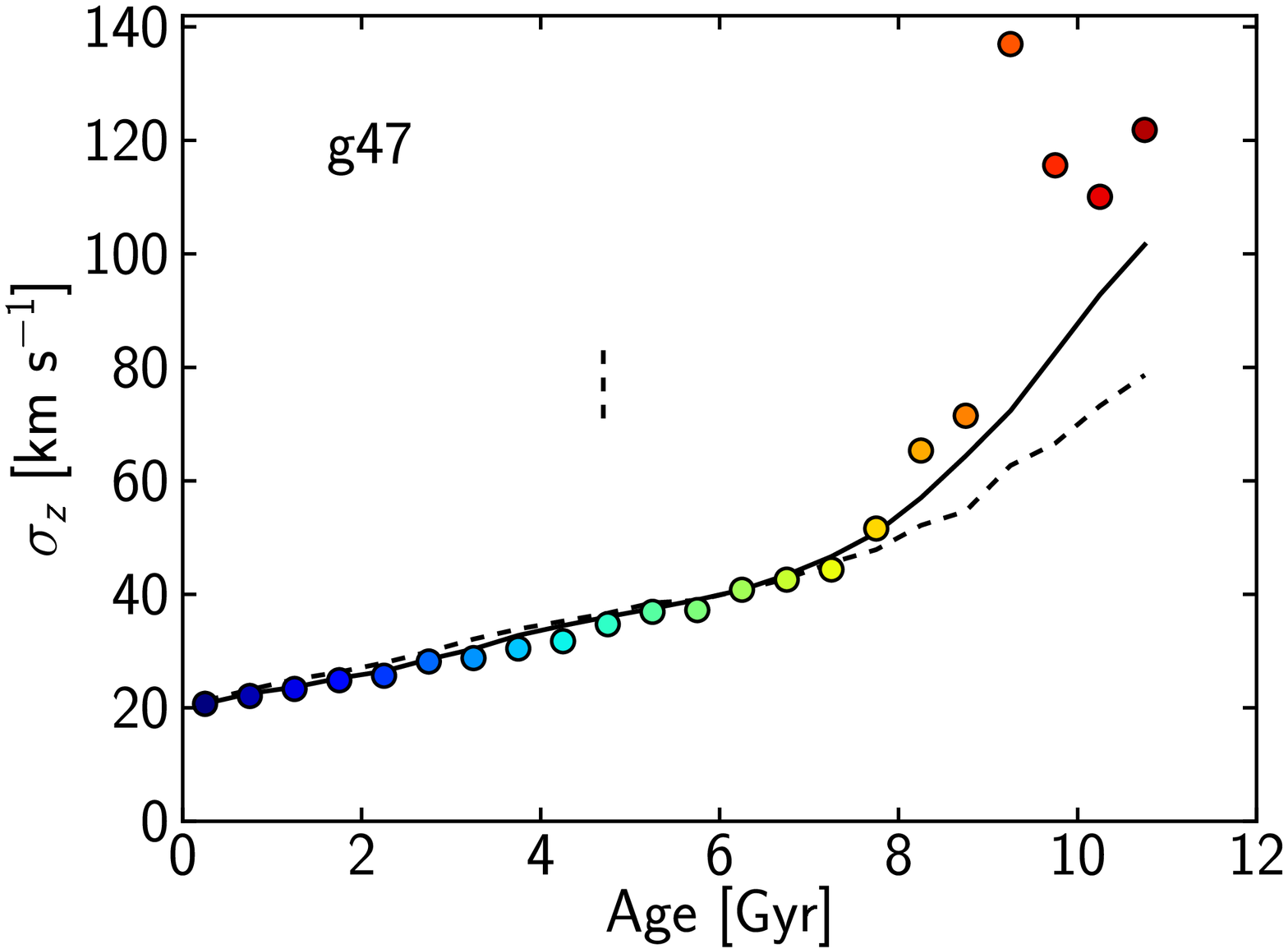}
\includegraphics[width=0.245\textwidth]{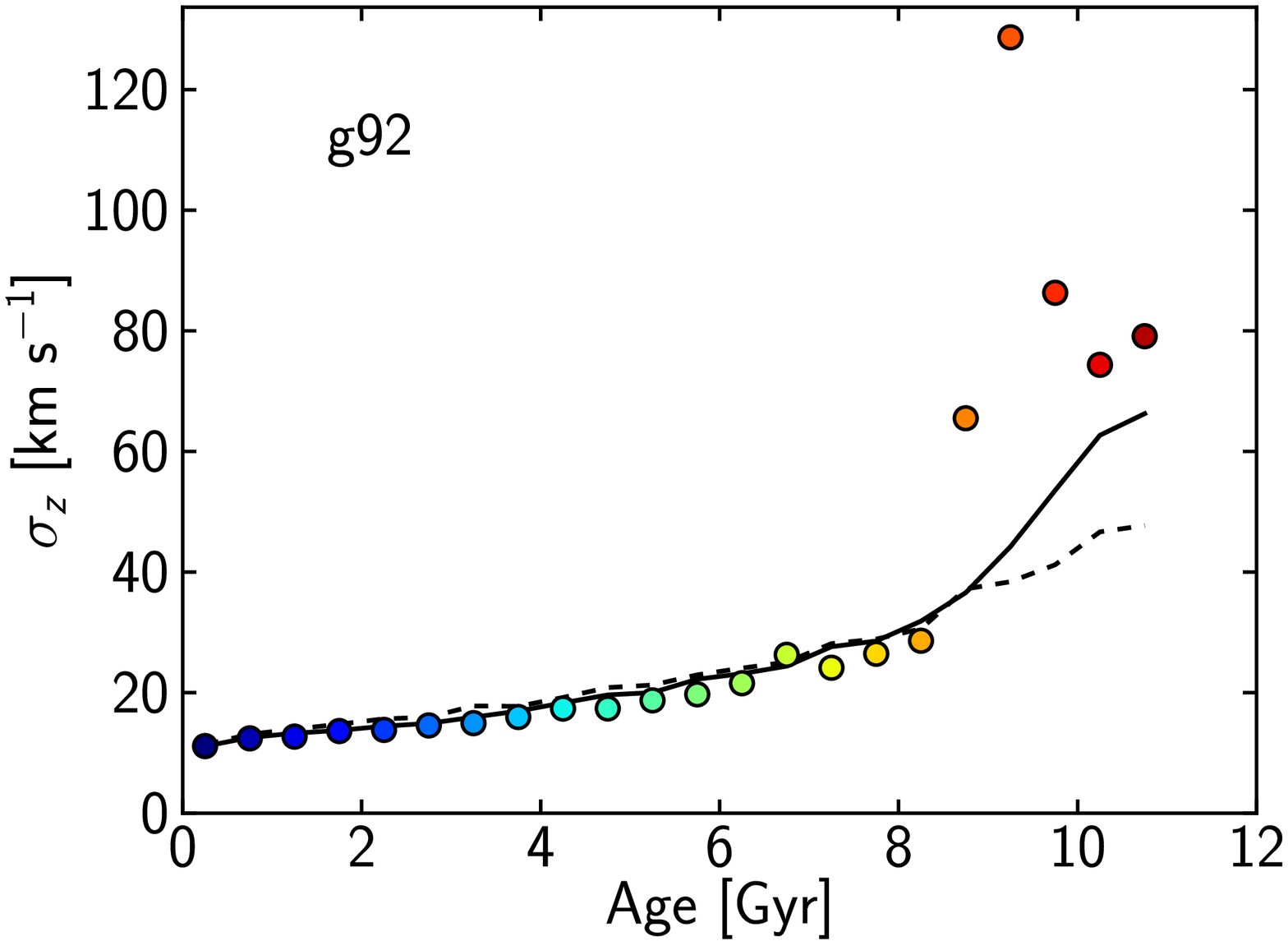}
\includegraphics[width=0.245\textwidth]{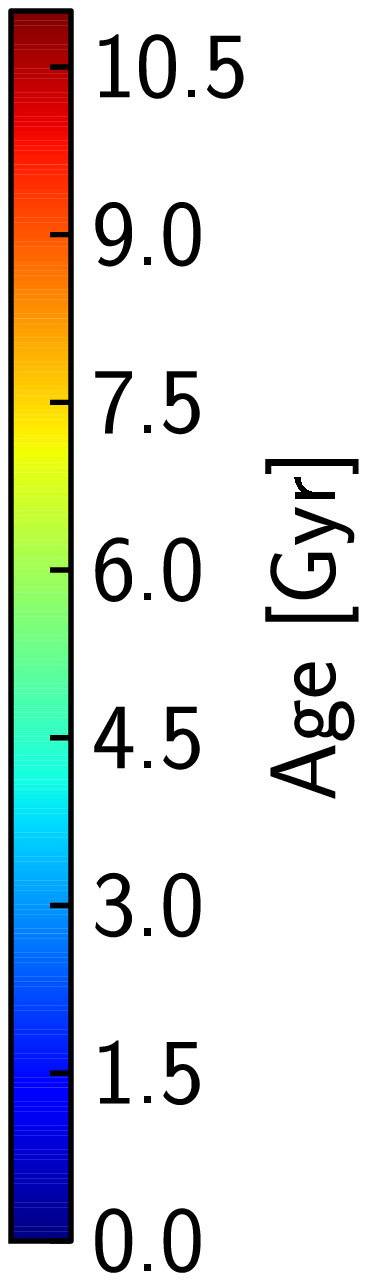}
\includegraphics[width=0.245\textwidth]{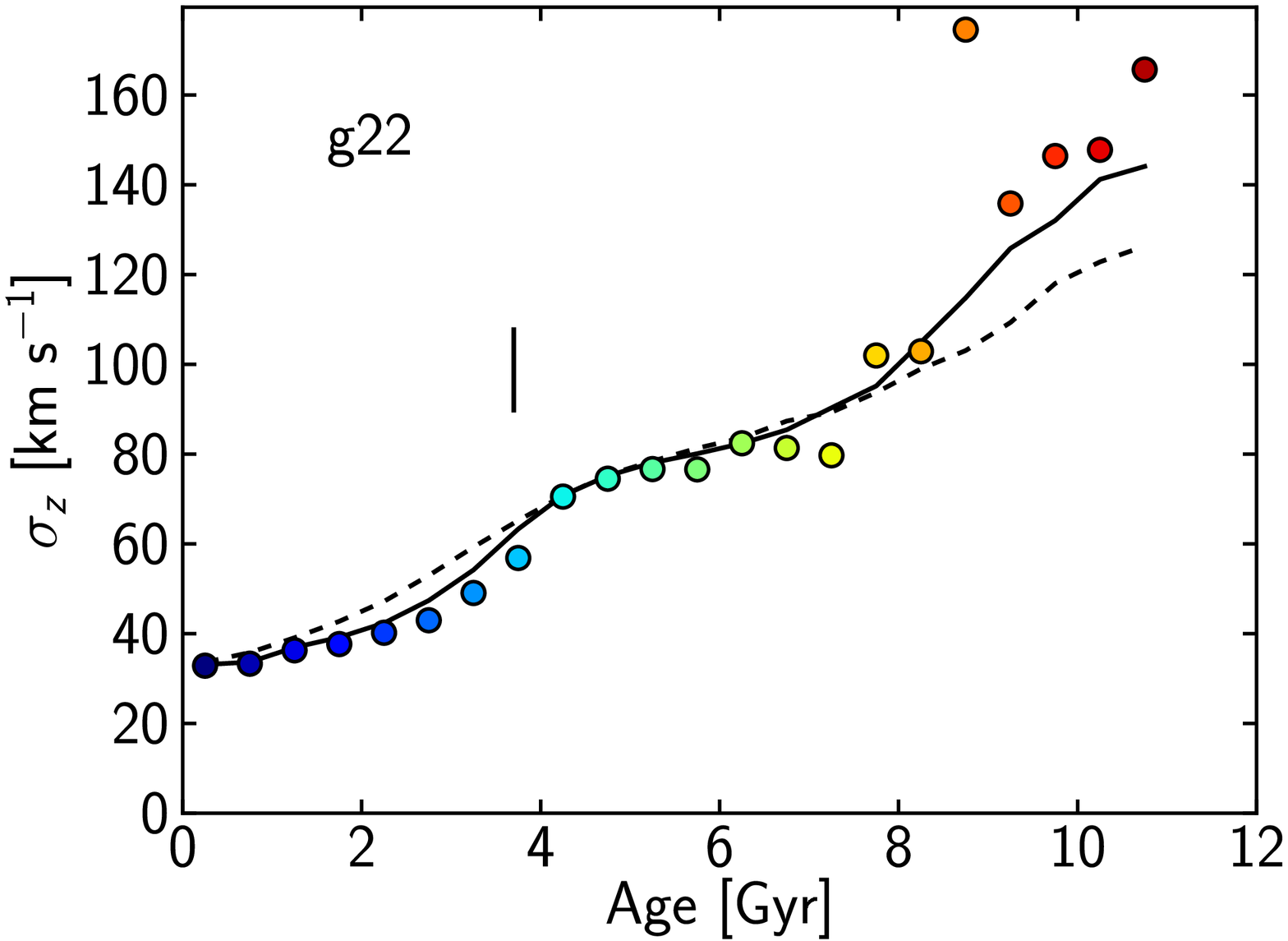}
\includegraphics[width=0.245\textwidth]{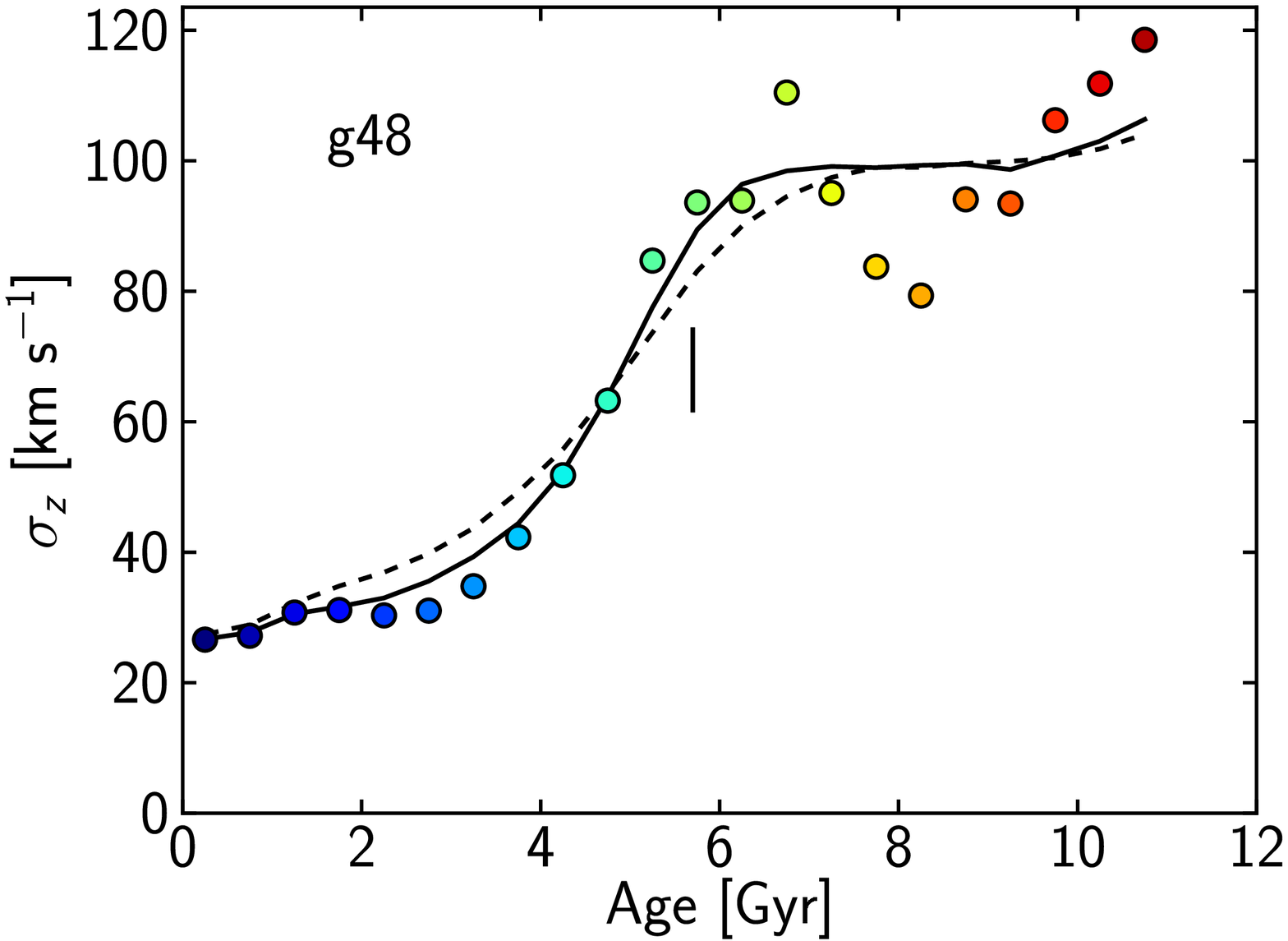}
\includegraphics[width=0.245\textwidth]{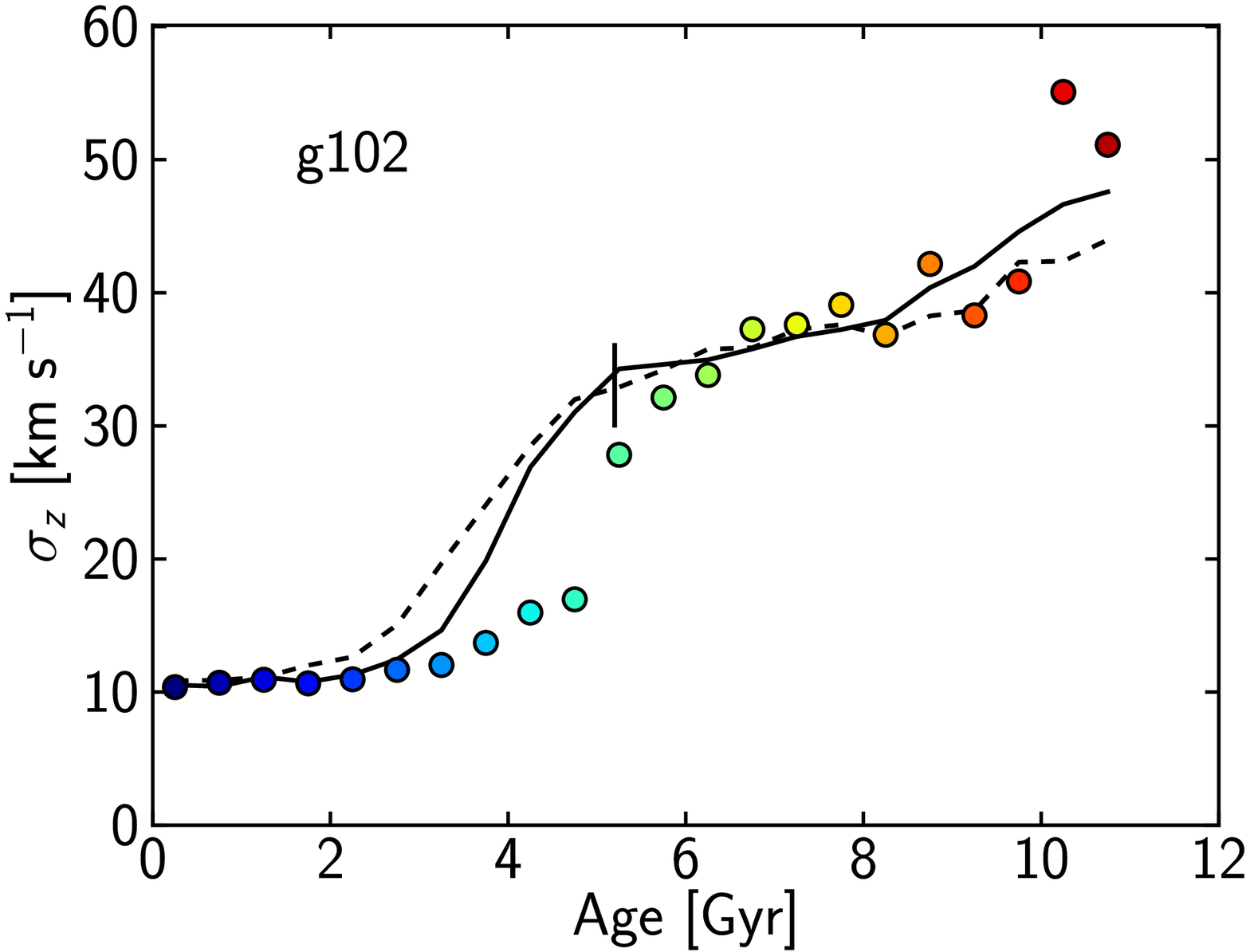}
\includegraphics[width=0.245\textwidth]{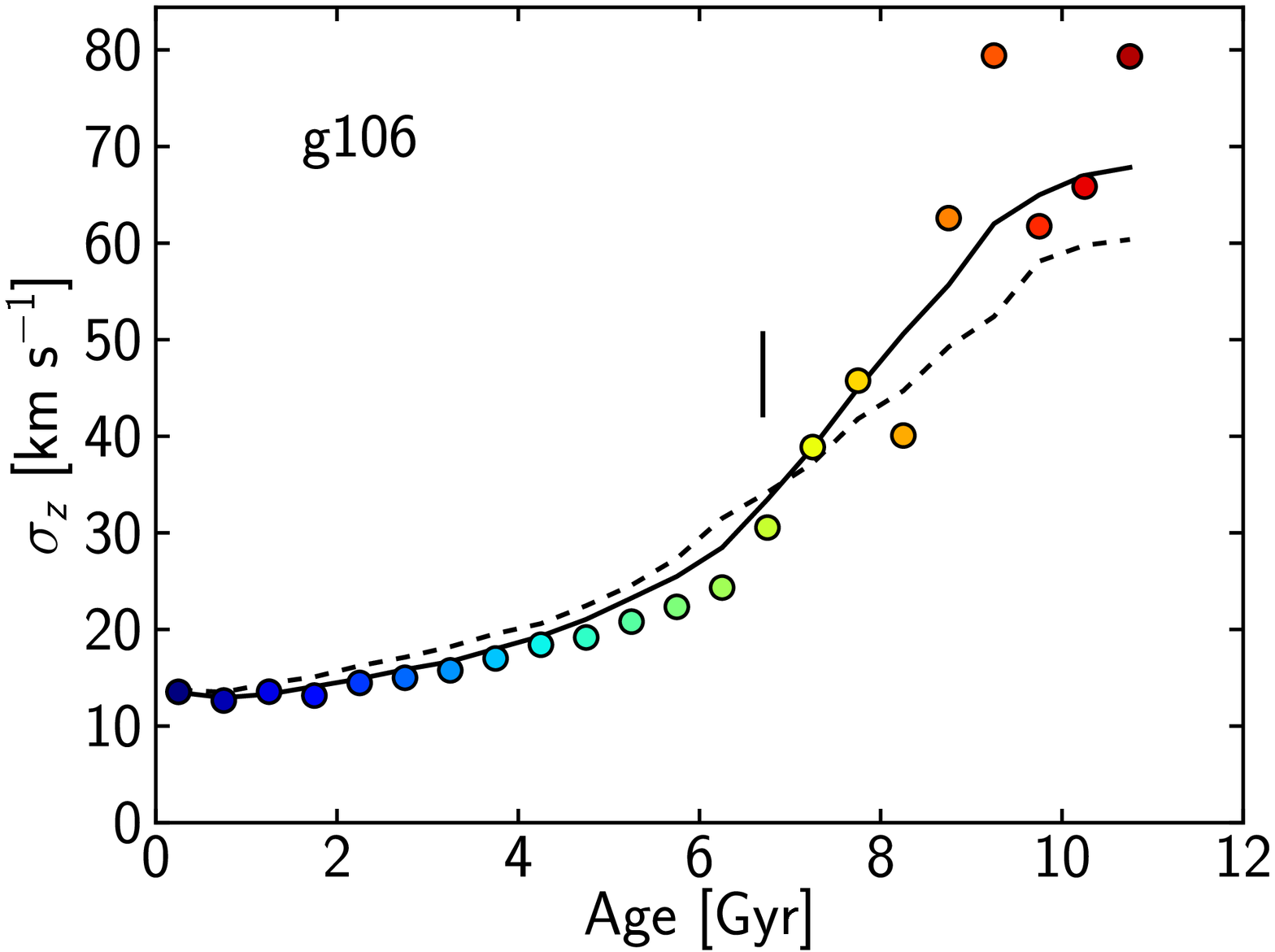}
\caption{Vertical velocity dispersion as a function of age at $2R_d$. The dots represent the actual value of \sz, while the lines correspond to \sz measured once a gaussian error of 20\% (solid line) or 30\% (dashed line) has been first added to the age of each particle. Galaxies in the top panels have quiescent merger histories for $z<1.5$, while galaxies in the bottom panels have more active histories. The small vertical lines mark the time of coalescence for the last merger undergone by each galaxy  for $z<1.5$ (the dashed line for g47 marks the end of a fly-by). We find a smoothly rising age-velocity relation for the quiescent galaxies, while mergers imprint jumps in the relation. The age errors do not change the shape of the AVR for quiescent galaxies, but smooth jumps for galaxies with mergers.}
\label{fig:AVR}
\end{figure*}

The simulations we use in this paper are the same as in Paper I, to which we refer the reader for a detailed description of the simulation technique. The sample consists of seven galaxies simulated using a zoom-in cosmological re-simulation technique \citep[see][]{Martig2009,Martig2012}. This technique consists in first running a ``dark matter only'' cosmological simulation, in extracting merger and accretion histories for a set of haloes, and in re-simulating these histories at higher resolution, starting at $z=5$ with a seed galaxy. We follow the evolution of that seed galaxy down to $z=0$ by including mergers as well as accretion of gas and dark matter as prescribed by the cosmological simulation.

The re-simulations use a Particle-Mesh code with a grid size of 150 pc, where gas dynamics is modelled with a sticky-particle algorithm.  The mass resolution is  $1.5\times 10^4$~M$_{\sun}$ for gas particles, $7.5\times 10^4$~M$_{\sun}$ for star particles ($1.5\times 10^4$~M$_{\sun}$ for star particles formed from gas during the simulation), and  $3\times 10^5$~M$_{\sun}$ for dark matter particles. For the galaxies studied here, there are typically  2 to 6 million star particles in the discs at $z=0$.

We model star formation following a Schmidt law with an exponent of 1.5 above a volume density threshold of 0.03 M$_{\sun}$pc$^{-3}$ (i.e., 1 H cm$^{-3}$). We take into account stellar mass loss \citep[see][]{Martig2010}, as well as kinetic feedback from supernovae explosions, where 20\% of the energy of the supernovae is distributed to neighbouring gas particles.

The seven chosen galaxies have a total stellar mass at $z=0$ of between $3\times 10^{10}$ M$_{\sun}$ and $2\times 10^{11}$ M$_{\sun}$, and bulge-to-total ratios between 0.04 and 0.5 (see Table 1 in Paper I). They were selected to have a range of formation histories as follows:

\begin{description}
\item \textbf{g37} and \textbf{g92} have quiescent histories: over their last 9 Gyr of evolution they only have interactions with mass ratios smaller than 1:50. 
\item \textbf{g47} is also mostly quiescent, but undergoes a 1:15 interaction with a satellite at $t \sim$8.5--9 Gyr, the satellite does not merge with g47 but significantly disturbs its morphology
\item \textbf{g22}  undergoes a 1:10 merger at $t=10$ Gyr
\item \textbf{g48} undergoes a 1:4 merger at $t=8$ Gyr
\item \textbf{g102} undergoes a 1:15 merger at $t=8.5$ Gyr
\item \textbf{g106} undergoes a 1:5 merger at $t=7$ Gyr
\end{description}

The properties of each galaxy are analysed at $z=0$. In this paper, we present the age-velocity relation for particles in age bins of 500 Myr. We first remove counter-rotating stars to minimize the contamination of our sample by halo stars. Then, we measure the vertical velocity dispersion for each population  within a 2 kpc-wide annulus centered at 2 R$_d$ (R$_d$ is the scale-length obtained from an exponential fit to the total stellar mass distribution), and up to a height including 95\% of stars of that population (limited to 5 kpc above the disc plane). We estimate \sz from the second moment of the velocity distribution; the resulting AVR is shown in Figure \ref{fig:AVR} for each of the seven galaxies.

Since errors on ages are a significant observational issue, we perform an additional test: we add a Gaussian error on the age of each stellar particle before re-computing the AVR using the blurred ages. We test  assumed age errors of 20\% and 30\%, which are typical of current observational measurements \citep{Holmberg2007,Soderblom2010}. The resulting AVRs are shown as black lines in Figure \ref{fig:AVR}.

In the next Section, we compare the overall shape of the AVR for quiescent and active galaxies, and discuss the values of \sz for young stars.

\section{The shape of the age-velocity relation}
The shape of the AVR is very different for galaxies with different merger histories (Figure \ref{fig:AVR}). Quiescent galaxies show a smooth rise of the AVR, while mergers leave well defined jumps. An exception is g106: it undergoes a 1:5 merger, but this merger happens early enough so that it leaves little trace in the AVR. In the following, we thus study g106 together with the quiescent galaxies.
\subsection{Quiescent galaxies}

Apart from the oldest populations, which have a high \sz in all cases, the quiescent galaxies show a very regularly rising \sz with age. Interestingly, the fly-by undergone by g47 does not leave any trace in the AVR at a radius of 2$R_d$, only leaving signatures in the outer disc of g47.

In the Milky Way, the AVR is often fitted by a power law corresponding to \sz $\propto t^{\alpha}$ in intermediate ages range (excluding young unrelaxed stars, and hot old stars). The heating index $\alpha$ is difficult to measure but usually found close to 0.5 \citep{Holmberg2007,Seabroke2007}, and its value has been used to discriminate between disc heating mechanisms \citep[e.g.,][]{Hanninen2002}. We similarly fit the simulated AVRs with a power law for ages between 2 and 7.5 Gyr (6.5 Gyr for g106), and find heating indices close to 0.5 (see Table \ref{tab-heating}). This is a good match to the AVR in the Milky Way, suggesting that the heating in our simulations is reasonable.

\begin{table}
 \caption{Heating index for the four most quiescent galaxies. The heating index is measured as the slope of a linear fit to the logarithm of the AVR for ages between 2 and 7.5 Gyr (6.5 Gyr for g106). We  show both the actual value of the index, and the value measured on the AVR when a 20\% or 30 \% age error is first taken into account (the black lines in Figure \ref{fig:AVR}).}
 \label{tab-heating}
 \begin{tabular}{@{}lccc}
  \hline
  Name & Actual value& With 20\% age error & With 30 \% age error\\
  \hline
 g37 &0.53 $\pm$  0.02&   0.52 $\pm$ 0.02 & 0.46 $\pm$ 0.02 \\
 g47 & 0.47 $\pm$ 0.03  & 0.46 $\pm$ 0.02& 0.39 $\pm$ 0.01 \\
 g92 & 0.52 $\pm$ 0.06 &  0.54 $\pm$ 0.04&0.49 $\pm$ 0.04\\
g106 &0.51 $\pm$ 0.04 & 0.62 $\pm$ 0.06& 0.61 $\pm$ 0.07 \\
  \hline
 \end{tabular}
\end{table}

\begin{figure}
\centering 
\includegraphics[width=0.45\textwidth]{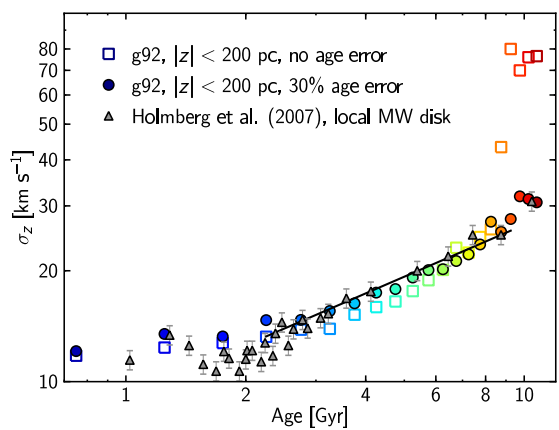}
\caption{Age-velocity relation for g92, compared with Milky Way data from Holmberg et al. (2007 -- grey triangles).  We also show a power law fit to the simulated AVR, the resulting exponent is 0.47$\pm$0.03 for g92, showing a good match to the observed slope. We selected only stars close to the disc plane to replicate the selection criterion of the Geneva-Copenhagen Survey. We find that this results only in a small overall number of old stars in the sample, and that, when age errors are added, the jump in \sz found for old stars is totally washed out, making the simulation consistent with GCS data.}
\label{fig:sigma_MW}
\end{figure}

When we add an error of 20\% or 30\% to the age of each stellar particle, for ``blurred ages'' smaller than 8 Gyr, we do not find significant changes to the shape of the AVR (black lines on Figure \ref{fig:AVR}). This is because the number density in this age range generally does not vary much and \sz increases smoothly with age. The heating indices are mostly unchanged with 20\% age errors, and are slightly reduced with 30 \% age errors, which means that the AVR is slightly flattened (see Table \ref{tab-heating}). An exception is g106, for which the stars with a true age of $\sim$ 8 Gyr (hotter because of the 1:5 merger), are now mixed also with younger stars so that the AVR is overall steeper, even when only 20\% age errors are included: the heating index is increased from 0.51  $\pm$ 0.04 to 0.62 $\pm$ 0.06. With such large error bars, though, g106 might still be consistent with the Milky Way.

In all galaxies, for the oldest stars,  age errors introduce significant changes to the AVR: \sz is lowered because the truly hot and old stars are now mixed with younger and colder stars, and because these young stars are more numerous. With age errors, the jumps in \sz seen for the oldest stars are then considerably smoothened, and might be missed by surveys.

To illustrate that point, we show in Figure \ref{fig:sigma_MW} a comparison between the AVR in the solar neighbourhood from the Geneva-Copenhagen Survey \citep[GCS,][]{Nordstrom2004,Holmberg2007} to one of our simulated galaxies. The observed AVR shows a smooth increase of \sz with age, with no jump for old stars, and a \sz of about 30 \kms for the oldest stars. This observation might seem in contradiction with other studies finding  \sz$\sim 45$ \kms for the oldest stars \citep[e.g.,][]{Quillen2001}, or similarly high values for populations with high [$\alpha$/Fe] ratios \citep{Lee2011,Bovy2012c,Haywood2013}. However, the GCS sample is limited to a small volume around the Sun, and contains very few metal-poor stars (typically [Fe/H] $< -0.7$), which are the stars for which \cite{Bovy2012c} find \sz $>$ 40 \kms. In addition, when selecting only stars close to the disc mid-plane ($|z| < 200$ pc) in our g92 simulation, we show in Figure \ref{fig:sigma_MW} that even if a jump in \sz is present for old stars, this jump is totally erased when 30\% age errors are included, and the simulated galaxy matches the observed AVR very well. In this simulation, we find that the absence of a jump is both the consequence of the age errors and of the selection of a sample close to the disc plane, dominated by kinematically cold, thin disc stars (in Figure \ref{fig:AVR}, which includes all stars of a given age, adding age errors is not enough to totally erase the jump).

This shows that in this sense, our simulations are consistent with observations of the solar neighbourhood, and that even if a jump in \sz were present in the Milky Way for old stars, it might not be detected by surveys limited to a small volume and with large age errors.

\subsection{The effect of mergers}

By contrast, the galaxies undergoing a merger show jumps in their AVRs, that appear at ages corresponding to the end of the merger. This is clearly seen for g22 and g102. For g48, where the merger mass ratio is about twice higher than in g22, the jump is not so sharp, and a lot of time is needed for the gas to cool down again: it takes about 2 Gyrs for the velocity dispersion to decrease after the galaxies merge.

It is interesting that even the small 1:10--1:15 mergers leave a well defined signature on the AVR (Figure \ref{fig:AVR}, bottom row). We also note here that mergers affect disc kinematics at all radii (as also found by \citealp{Velazquez1999}), while the fly-by in g47 only affects outer regions. Whether this is a general feature of mergers vs fly-bys would need to be verified with larger samples of simulations.

These results are consistent with those of \cite{House2011}, where mergers were similarly found to create jumps in the AVR, and are also consistent with  idealized studies of the effect of mergers on disc kinematics \citep[e.g.,][]{Quinn1993,Velazquez1999,Kazantzidis2009}. 

However, contrary to quiescent galaxies, blurring the ages has a greater effect on the shape of the AVR for galaxies with mergers (black lines in Figure \ref{fig:AVR}). The jumps are considerably smoothened, so that they do not trace exactly the end of mergers anymore (especially for 30\% errors). Note also that for g22 the jumps actually becomes far less noticeable for age errors of 30\%, which suggests that errors on ages of 20\% or less are critical to detecting the traces of minor mergers in the observed AVR.

\subsection{The velocity dispersion of young stars}
\begin{figure}
\centering 
\includegraphics[width=0.45\textwidth]{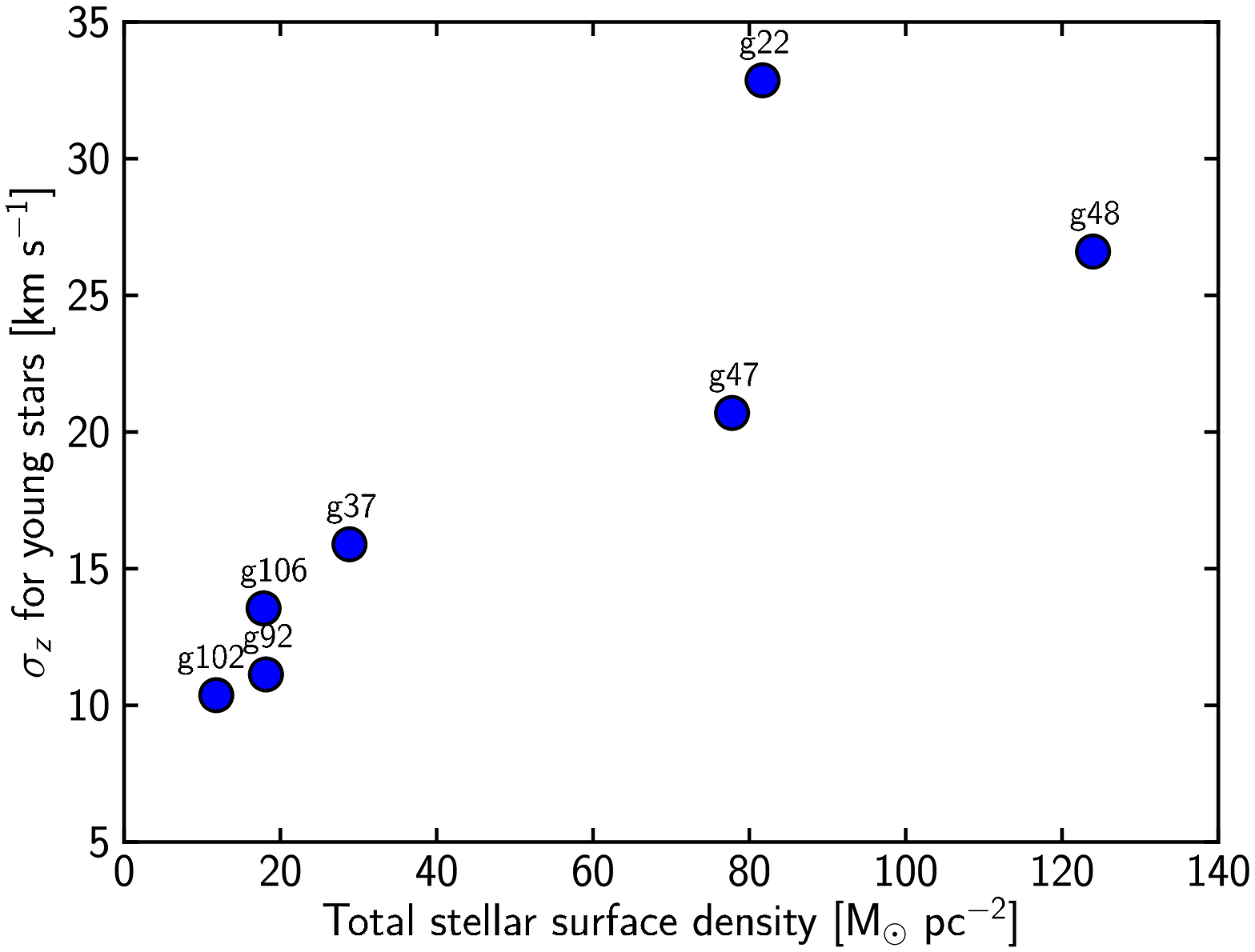}
\includegraphics[width=0.45\textwidth]{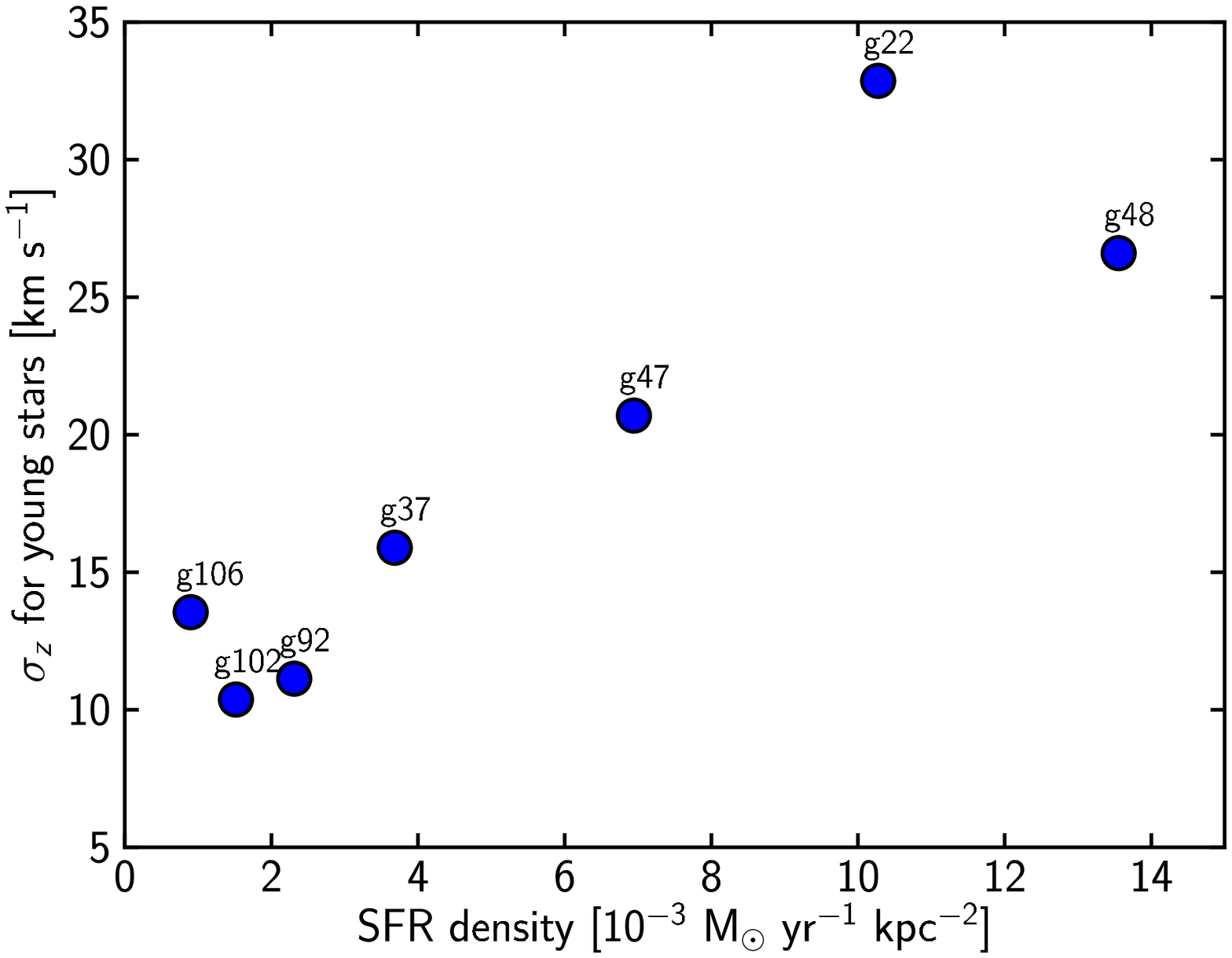}
\caption{Relation between the vertical velocity dispersion \sz for young stars at 2$R_d$ and the stellar surface density at that same radius (top panel) as well as the local star formation rate surface density (bottom panel) for all seven galaxies. Both \sz and the SFR are defined as corresponding to the last 500 Myr of evolution before $z=0$.}
\label{fig:sigma_young}
\end{figure}
For our seven galaxies, we find that the youngest stars (age $<$ 500 Myr) have a vertical velocity dispersion that varies between 10 and 33 \kms at 2 $R_d$. In all galaxies, this value declines at increasing galactocentric distances, as shown in Paper I (Figure 6). For instance at 3 $R_d$, \sz for young stars varies between 7 and 23 \kms depending on the galaxy.

It is likely that stars in our simulations are born too kinematically hot, because we do not fully resolve the structure of the inter-stellar medium. In the solar neighbourhood, \cite{Aumer2009} find that the bluest (i.e., youngest) stars have a \sz of only 6 $\pm$ 1 \kms. This value is similar to the velocity dispersion of molecular clouds: \cite{Stark1989} measured a cloud-cloud velocity dispersion of 7.8 $\pm$ 0.6 \kms for clouds within 3~kpc of the Sun. This dispersion depends on cloud mass, and the most massive GMCs have a cloud-cloud velocity dispersion of only 4 \kms \citep{Stark2005,Stark2006}.
If stars are born with such low velocity dispersions, they are then heated to  \sz $\sim$ 10--15 \kms within a few hundred million years (see also Figure \ref{fig:observations}). This heating can be very fast: in an idealized simulation of a Milky-Way type galaxy evolving in isolation, \cite{Renaud2013} find that stars initially born with a total velocity dispersion of 10 \kms are heated to 15 \kms in only 10 Myr. Assuming an isotropic velocity dispersion, this translates into a \sz at birth of 5.8 \kms. To obtain such a cold stellar component, they used a sub-parsec resolution (down to 0.05~pc), that we cannot achieve in cosmological simulations (our resolution is 150 pc).

It seems however that even if stars are not formed very cold, and even if we do not resolve that phase of intense early heating, the values of \sz we obtain at 0.5--1~Gyr are reasonable compared to the Milky Way: in four of our simulations, young stars have a \sz between 10 and 17 \kms.
 
Our three other galaxies (g47, g48, and g22) have a higher \sz for their young stars, up to 33 \kms for g22. These reflect a higher gas velocity dispersion for these galaxies. As shown in Figure \ref{fig:sigma_young}, what they have in common is both a higher stellar surface density, as well as higher star formation rate surface density, compared to the kinematically colder galaxies.

A higher surface density has been shown to increase gas velocity dispersion through gravitational instabilities  \citep{Wada2002, Agertz2009, Bournaud2010}. Another source of turbulence is the feedback from supernovae explosions, that increases with the star formation rate. \cite{Dib2006} showed that the atomic gas velocity dispersion does not depend on the SFR surface density up to a threshold corresponding to a starburst regime: for SFR densities greater than 0.5--1$\times 10^{-2}$\msun yr$^{-1}$ kpc$^{-2}$, the atomic gas velocity dispersion increases with the SFR density \citep[see also][]{Agertz2009, Tamburro2009}. Interestingly, this threshold value is similar to the SFR density we find in our kinematically hottest galaxies, suggesting that feedback could be the source of the high velocity dispersion in the gas disc of these galaxies. Finally, it is also possible part of the heating is provided by satellite galaxies, either merging with the discs, or simply orbiting around the main galaxies.

It has also been observed that for molecular gas, the cloud-cloud velocity dispersion decreases with radius, and differs among local galaxies \citep{Wilson2011}; it is roughly twice smaller than the atomic gas velocity dispersion \citep{Tamburro2009}. Plausibly, different galaxies could have a different \sz for their young stars.

In simulations, though, the gas velocity dispersion is also the result of numerical effects \citep[e.g.,][]{Pilkington2011, Agertz2013}.
\cite{House2011} compared the vertical velocity dispersion in various simulations performed with different codes, resolutions and recipes. They find a ``floor'' in \sz of 15--20 \kms even in high resolution simulations (50 pc), except when the star formation threshold is increased from 0.1 cm$^{-3}$ to 100 cm$^{-3}$. 

Our simulations are difficult to directly compare to the above results, because \sz varies as a function of radius and from one galaxy to another. 
What we find is that an increase of the resolution by a factor of 2 does not change our results (see Section 6). This could mean that a much higher (sub-parsec) resolution would be necessary to successfully reproduce a very low \sz for stars at birth \citep[as in][]{Renaud2013}. We also find that increasing the star formation threshold from 1 to 5 H cm$^{-3}$ decreases the overall normalization of the AVR (because we did not increase the star formation efficiency, so that the total mass is lower), but does not change its shape, in particular there is no fast heating for young stars.

\section{The origin of the AVR}

\begin{figure}
\centering 
\includegraphics[width=0.45\textwidth]{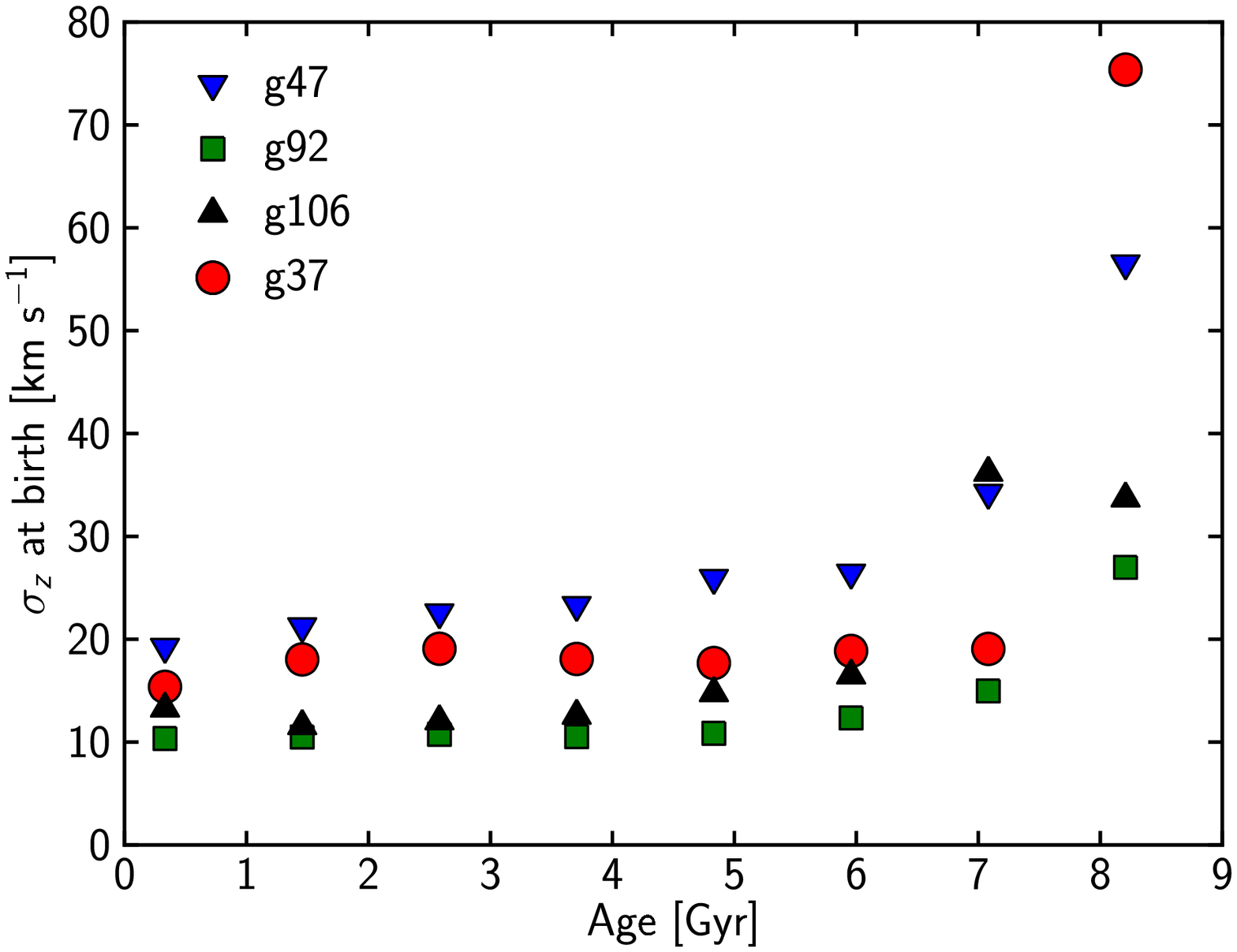}
\includegraphics[width=0.45\textwidth]{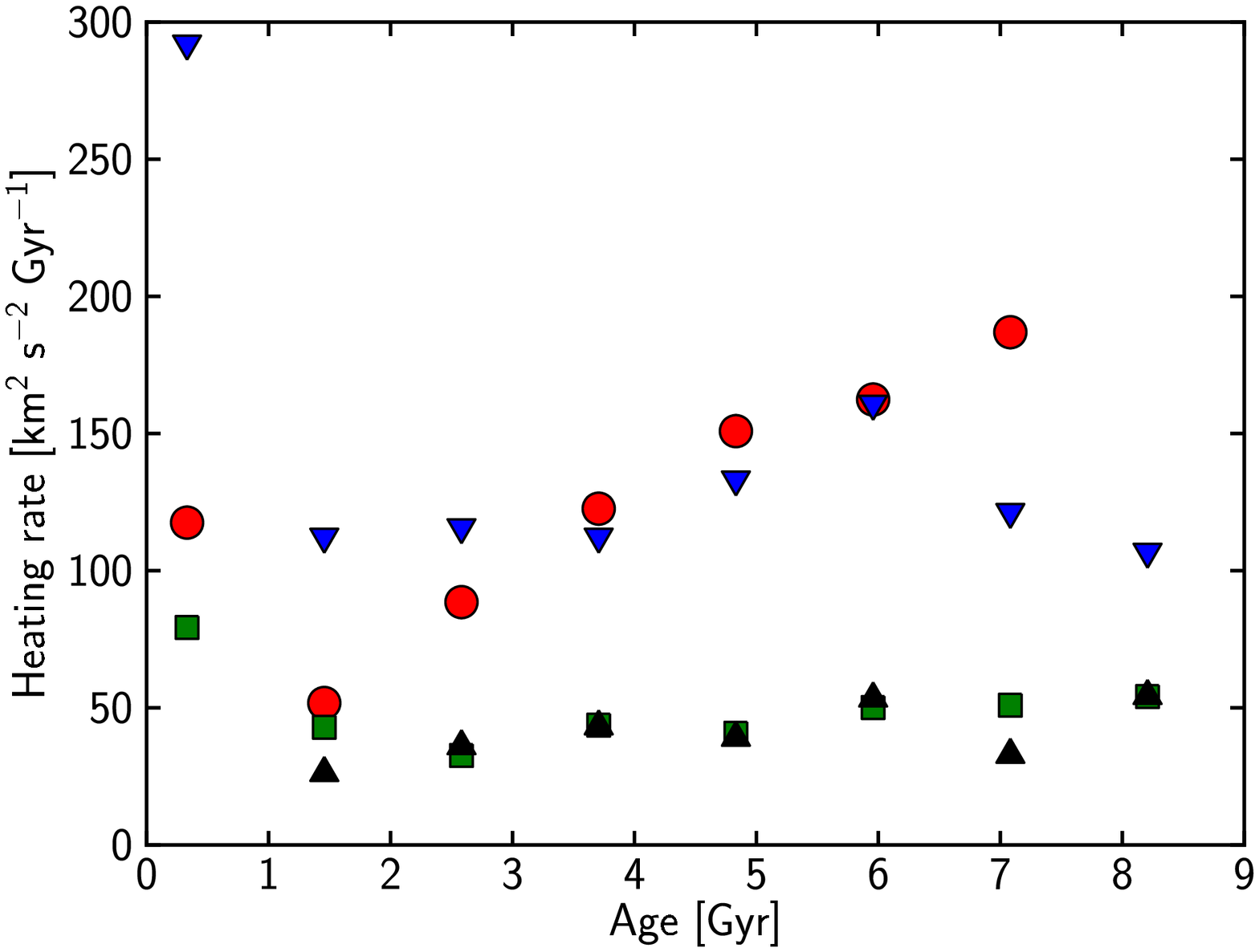}
\caption{Investigating the origin of the slope in the age-velocity relation. The top panel shows \sz at birth for a sample of mono-age populations in four of our simulated galaxies. The bottom panel shows the heating rate $D$ for the same populations, where $D$ is defined by  $\sigma_{z}^2(t)=\sigma_{z,0}^2+D (t - t_0)$. While the oldest populations are born hot, for stars younger than $\sim 7$ Gyr the \sz at birth is relatively constant with time, and the slope of the AVR is due to subsequent heating of the stars.}
\label{fig:sigma_birth}
\end{figure}

To understand the origin of the continuous increase of \sz with age in quiescent galaxies, we study how \sz varies with time for each mono-age population. In the top panel of Figure \ref{fig:sigma_birth} we show \sz at birth for a sample of mono-age populations for the three quiescent galaxies, as well as for g106, which shows similar characteristics to these galaxies (we here select stars being at 2$R_d$ at $z=0$, and trace them back to their time and place of birth). We find in all cases that 8--9 Gyr old stars form a different population, and are born significantly hotter than younger stars. They trace the end of a violent, high redshift phase of star formation. For g47 and g106, this initial phase extends a bit longer, with stars still being born hot at ages of 7--8 Gyr. For instance, for g106, around 10\% of stars of that age and at that radius are accreted from satellites, and while the interactions contribute to disc heating, the total \sz for these hot populations is dominated by the high \sz at birth for stars born within the  galaxy itself.
After that phase of mergers, the younger stars are then generally born with a relatively constant \sz, even though a slight decrease of the birth \sz with time can usually be noticed (Figure \ref{fig:sigma_birth}).

We study how each mono-age population is heated by following the time evolution to $z=0$ of the vertical velocity dispersion of its stars (again, for stars being at 2$R_d$ at $z=0$). 

disc heating can be modelled as a diffusion process in velocity space, in which case we can write $\sigma_{z}^2(t)=\sigma_{z,0}^2+D (t - t_0)$ (with $t_0$ and $\sigma_{z,0}$ the values at birth). $D$ is the diffusion coefficient, or the heating rate, and is plotted on the bottom panel of Figure \ref{fig:sigma_birth} for a sample of mono-age populations (we simply measure it from the difference between $\sigma_{z}^2$ at birth and at $z=0$). 

These heating rates are generally below 200 km$^2$s$^{-2}$Gyr$^{-1}$. They differ between galaxies: the lowest values are achieved for g92 and g106 (with $D\sim 40$~km$^2$s$^{-2}$Gyr$^{-1}$ for most mono-age populations), while g47 and g37 undergo more heating ($D\sim$~100--150 km$^2$s$^{-2}$Gyr$^{-1}$). While part of the heating could be due to spurious numerical effects (see discussion in Section \ref{sec-resolution}), such numerical effects should be present in all simulations, still allowing comparisons between simulations. For instance the much larger heating rate found in g37 compared to g92 suggests that this larger heating has a physical and not numerical origin.

For quiescent galaxies the slope of the AVR in our simulations is thus governed by two separate effects: the oldest stars are born hot, while the rest of the disc (with ages under 8 Gyr) is mostly born with a similar \sz and subsequently heated. These effects combine to produce a continuously rising \sz with increasing age.
This differs from the picture presented in \cite{Bird2013}, where the vertical structure of the disc is mostly imprinted at birth (with also some subsequent heating). This could be the result of a different simulation technique (potentially for instance a stronger supernova feedback in their simulations), with young stars maybe born too hot so that there is not much room for subsequent heating.

\section{Investigation of heating mechanisms}
\subsection{Numerical effects}\label{sec-resolution}

In the simulations presented in this paper, the relatively high \sz for young stars, and the increase of \sz with age for stars in galaxies with quiescent merger histories could both be numerical artefacts. In the Milky Way, the exact shape of the AVR is debated, but some studies find a low \sz for young stars, and a rapid increase of \sz in the first Gyr. If young stars are too hot in the simulations, we might just simply miss that first phase of intense heating and just have slower, more regular heating.

The velocity dispersion of stars at birth can be affected by the resolution of the simulation, by the recipe for star formation (and in particular the value of the gas density threshold above which stars are formed), by the recipe for supernova feedback (affecting the gas velocity dispersion), and also by our gas dynamics algorithm itself. In turn, subsequent heating is probably only affected by the resolution.

Low resolution can result in significant two-body heating, leading to equipartition of energy  between rotational and random motion, thus thickening discs \citep[e.g.,][]{Quinn1993}. Massive dark matter particles are efficient at scattering stellar disc particles \citep{Governato2004,Mayer2004,Kaufmann2007}, and also at increasing the gas thermal energy \citep{Steinmetz1997}. The effects on gas should however be negligible for dark matter particle masses below $10^9$ \msun \citep{Steinmetz1997}, which is the case for all modern zoom-in cosmological simulations. The heating of the stellar disc should also be limited for halos containing a few million dark matter particles. For instance, \cite{Widrow2005} show discs undergoing only some minor heating in simulations using 2 million particles with each a mass of $2.9\times  10^5$ \msun (similar to the dark matter particle mass in our simulations). Note also that \cite{Hanninen2002} show that halo black holes with a mass of $10^6$ \msun are not efficient at heating discs. Compared to \cite{Hanninen2002}, our use of a Particle-Mesh technique should reduce heating even further \citep[e.g.,][]{Sellwood2013}.

For modern simulations, \cite{Sellwood2013} argues that the most important source of numerical heating is actually the interactions between disc particles themselves, not between disc and halo. This effect is strongly dependent on the number of particles in the disc, and  a few million disc particles are required to limit relaxation in simulated discs. While this result is not directly relevant for our simulations because we use a different simulation technique, the fact that all our simulated discs contain a few million stellar particles is encouraging.

Finally, in addition to relaxation effects, resolution also influences the growth of non-axisymmetric features in discs \citep{Kaufmann2007,Khoperskov2007}, and the properties of infalling satellites, both potential sources of heating. In the end, however, \cite{House2011} found that resolution did not have a strong impact on the AVR in their simulated galaxies, while increasing the star formation threshold from 0.1 to 100 H cm$^{-3}$ strongly decreased the velocity dispersion of young stars.

\begin{figure}
\centering 
\includegraphics[width=0.45\textwidth]{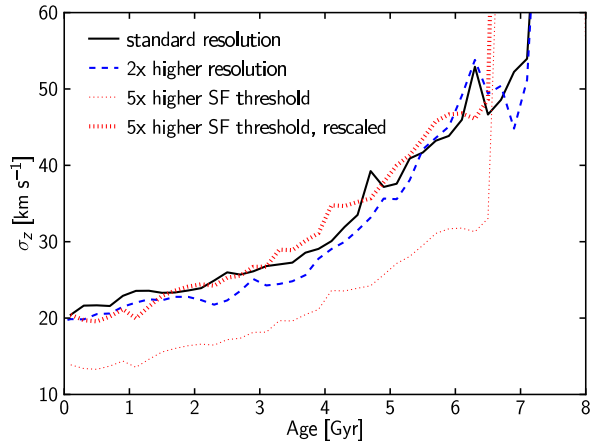}
\caption{Tests of the effect of an increased resolution (dashed blue line) or an increased star formation threshold (dotted red lines) on the shape of the age-velocity relation.}
\label{fig:resolution_test}
\end{figure}

We have tested the effect of resolution and star formation threshold on one of the simulated galaxies from the parent sample of \cite{Martig2012}. Part of the results from these tests were already presented in Appendix A of \cite{Martig2012}, where it was found that the increased resolution did not modify the stellar mass and radial density profile of that simulated galaxy. We here extend these tests to the AVR.

The resolution test consists in a twice higher spatial resolution (corresponding to 75 pc), and a mass resolution increased by a factor of 6 (the mass of a gas particle is then 2500 M$_{\odot}$). The test of the star formation threshold consists in increasing this threshold by a factor of 5 in the simulation at standard resolution (this moves the threshold from 1 to 5 H cm$^{-3}$).

A twice higher resolution only has a very small impact on the AVR (Figure \ref{fig:resolution_test}). We find that the \sz of young stars is similar to values at standard resolution, and the evolution of \sz with age also has the same shape. It just seems that \sz is consistently 1--2 \kms lower at high resolution, but this does not reconcile our simulations with observations of a fast rising AVR for small ages in the Milky Way. Note also that the constant offset of \sz values at higher resolution indicates that whatever the heating mechanisms are, they are not strongly affected by resolution (at least, for the resolutions tested here).

Increasing the threshold for star formation has a stronger effect on the simulated galaxy, partly because we did not increase the star formation efficiency so that the overall number of stars formed is lower, and the total stellar mass is only 80\% of the mass at standard resolution. With the higher threshold, the overall values of \sz are thus shifted to lower values. In particular, \sz for young stars becomes 14 \kms instead of 20 \kms. However, the shape of the AVR is not affected, in particular there is no fast heating from 0 to 1 Gyr. If we arbitrarily re-scale all the values of \sz to match the values of \sz for young stars in the standard simulation, we then find a good match between both simulations.

\subsection{Effect of radial migration}
\begin{figure*}
\centering 
\includegraphics[width=0.35\textwidth]{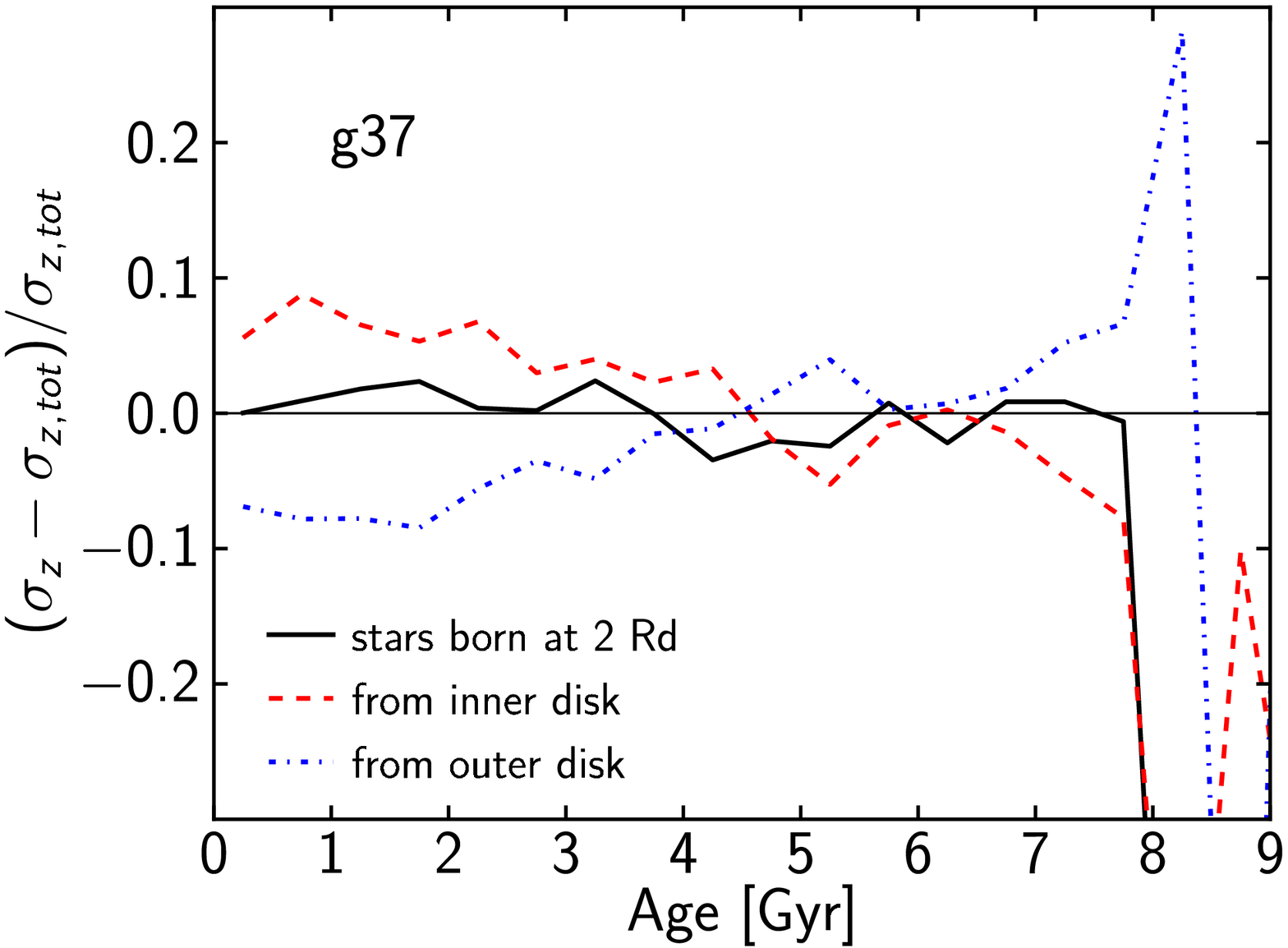}
\includegraphics[width=0.35\textwidth]{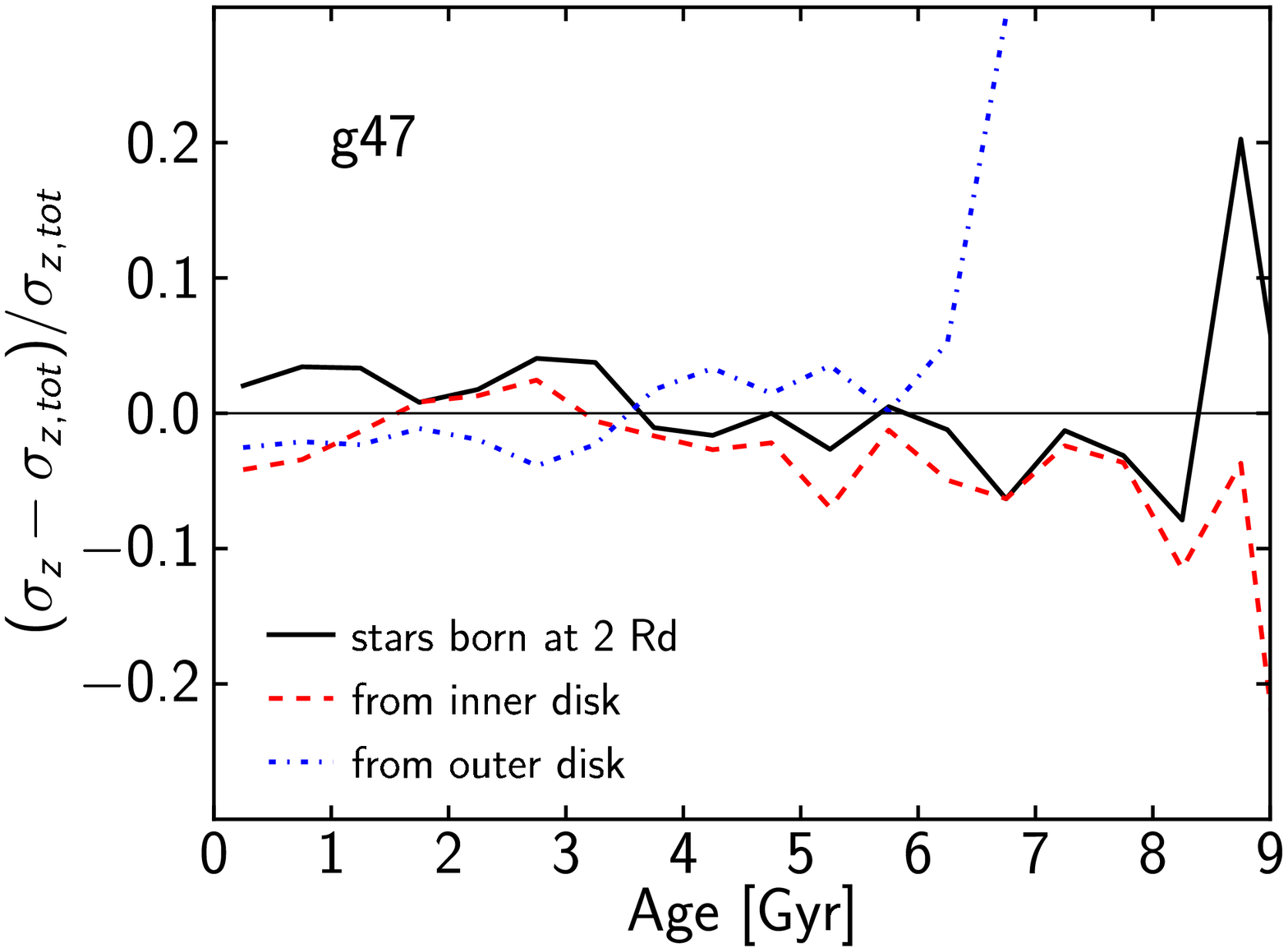}
\includegraphics[width=0.35\textwidth]{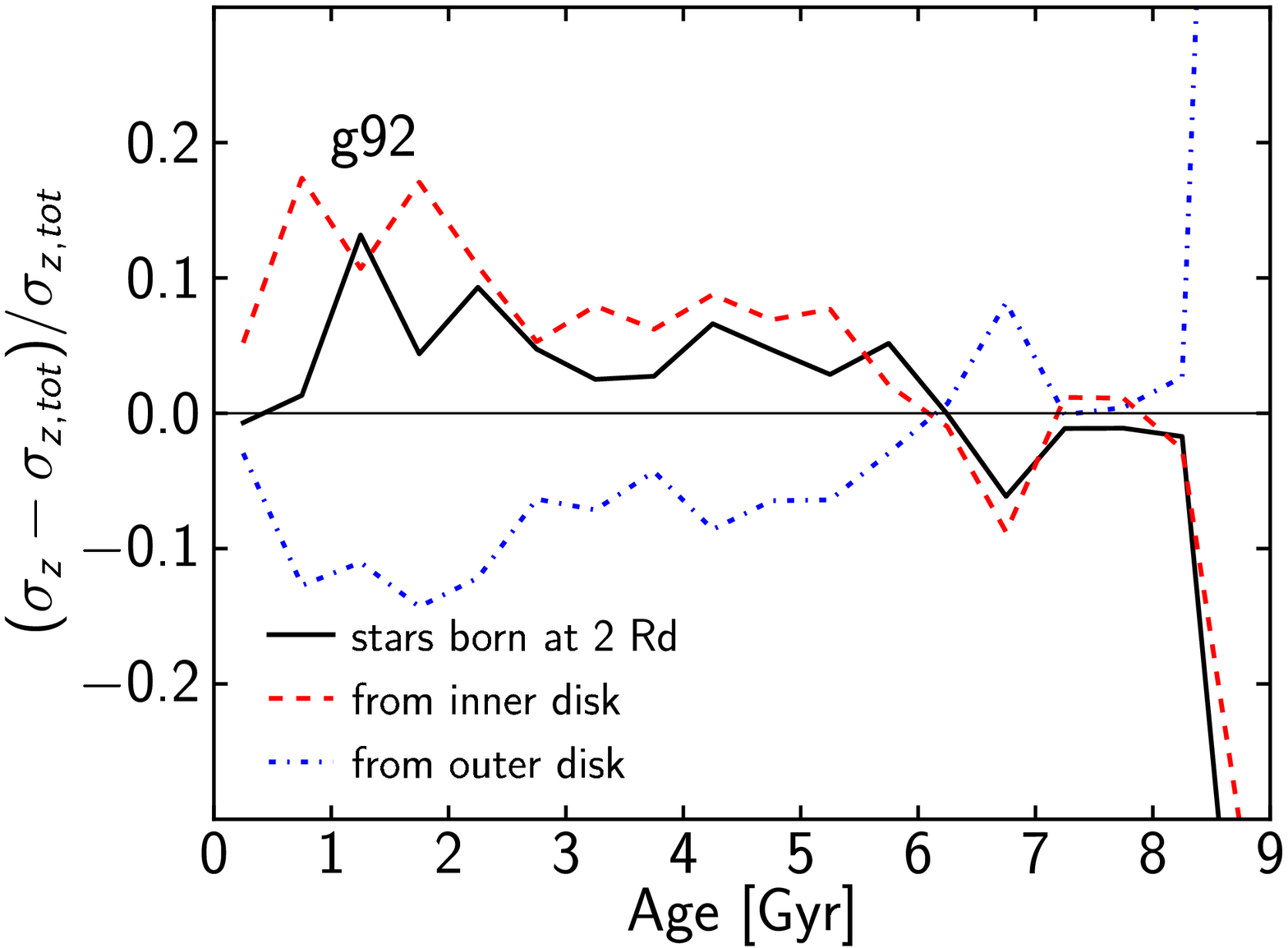}
\includegraphics[width=0.35\textwidth]{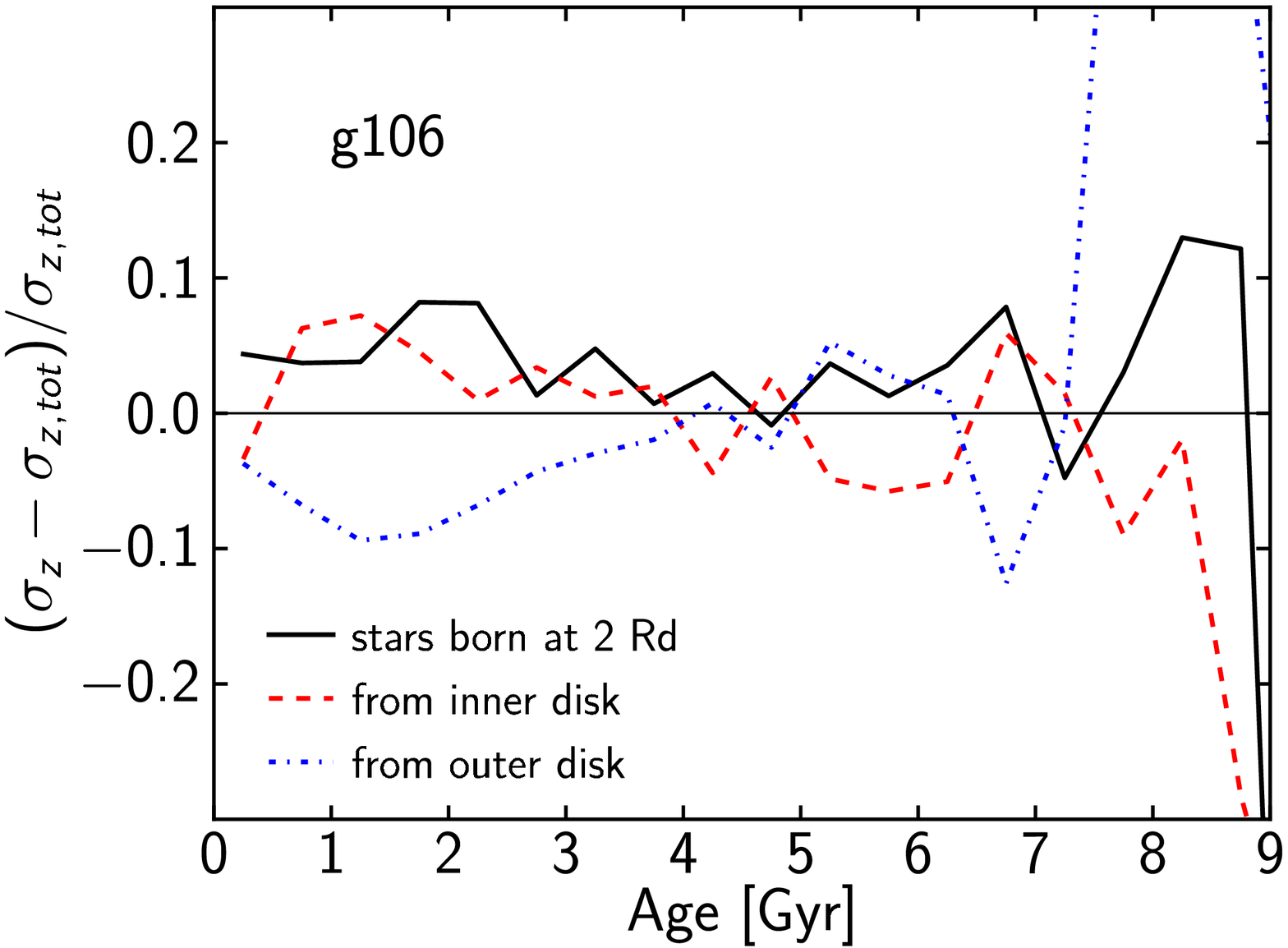}
\caption{Contribution of different populations of stars to the total vertical velocity dispersion as a function of age. For each mono-age population, we measure the fraction of its total \sz (measured at 2$R_d$ as previously) that is due to stars born in-situ (i.e. in a 2 kpc ring centered on 2$R_d$), stars born in the inner disc, and stars born in the outer disc. For stars younger than 7--8 Gyr, the total velocity dispersion is very similar to the dispersion for stars born in-situ, meaning that radial migration only has a minor effect on \sz at that radius (if anything, it slightly cools the disc because of the combined effect of inwards and outwards migrating stars). }
\label{fig:sigma_birthplace}
\end{figure*}
Recently, radial migration has been proposed as a source of disc heating \citep{Schonrich2009a,Schonrich2009b, Loebman2011, Roskar2013}, but we find no evidence for this in our simulations (as also discussed in \citealp{Minchev2012} and \citealp{Minchev2013}, and is consistent with \citealp{Bird2013}). In Figure \ref{fig:sigma_birthplace}, we show the fractional contribution of different types of stars (born in-situ, formed in the inner disc, formed in the outer disc) to the total \sz of each mono-age population. We find that for stars younger than 7--8 Gyr, the stars born in-situ mostly determine the total \sz (the total \sz only differs from the \sz of in-situ stars by less than 10\% in all cases). Similarly to the findings by Minchev et al. (2012, their Figure 5), stars coming from the inner disc tend to have a slightly larger \sz, and the opposite is true for stars from the outer disc. The net effect of migration is however to slightly cool the disc, because the heating from inner-disc stars is more than cancelled by the cooling from outer disc stars. Only in the very outer disc does the heating from outwards migrators dominate, creating some flaring \citep{Minchev2012}. We emphasize here the importance of considering both inwards and outwards migrators when studying the effect of migration on disc heating or thickening.

For the oldest stars, though, stars coming from the outer regions are the hottest. This is because the early mergers affect more significantly the outer disc, creating a flared morphology. The outwards migration of cold but old stars from the inner disc is shown by \cite{Minchev2014} to be a key ingredient in reproducing some chemo-dynamical features observed in the RAVE and SEGUE surveys (namely, the low vertical velocity dispersion for the most alpha-rich stars).

Ruling out migration as a source of heating at a radius of 2$R_d$ still leaves a lot of possible sources of heating in the simulated discs. We investigate these sources of heating in the following sections.

\subsection{Heating due to disc growth}
\begin{figure*}
\centering
\includegraphics[width=0.33\textwidth]{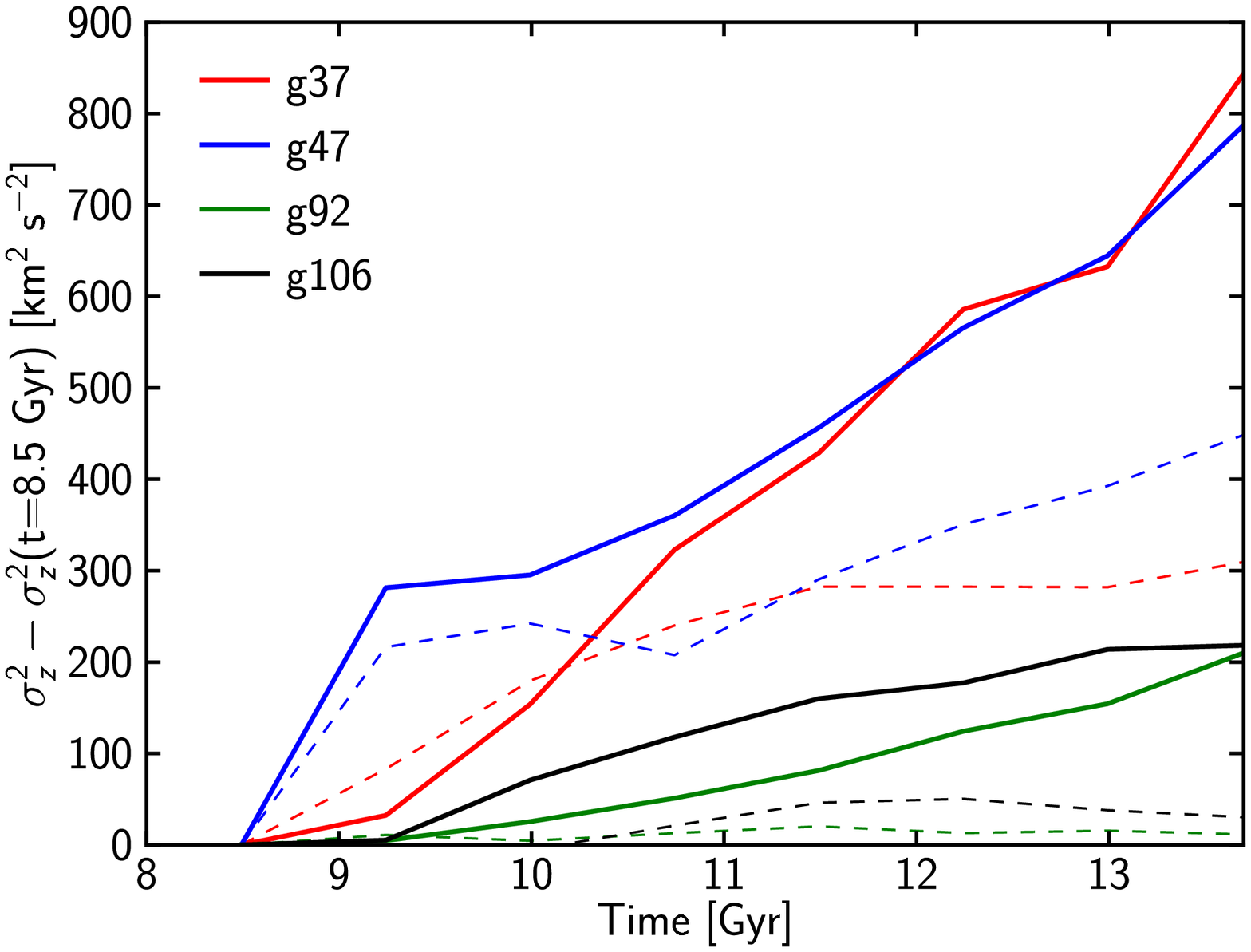}
\includegraphics[width=0.33\textwidth]{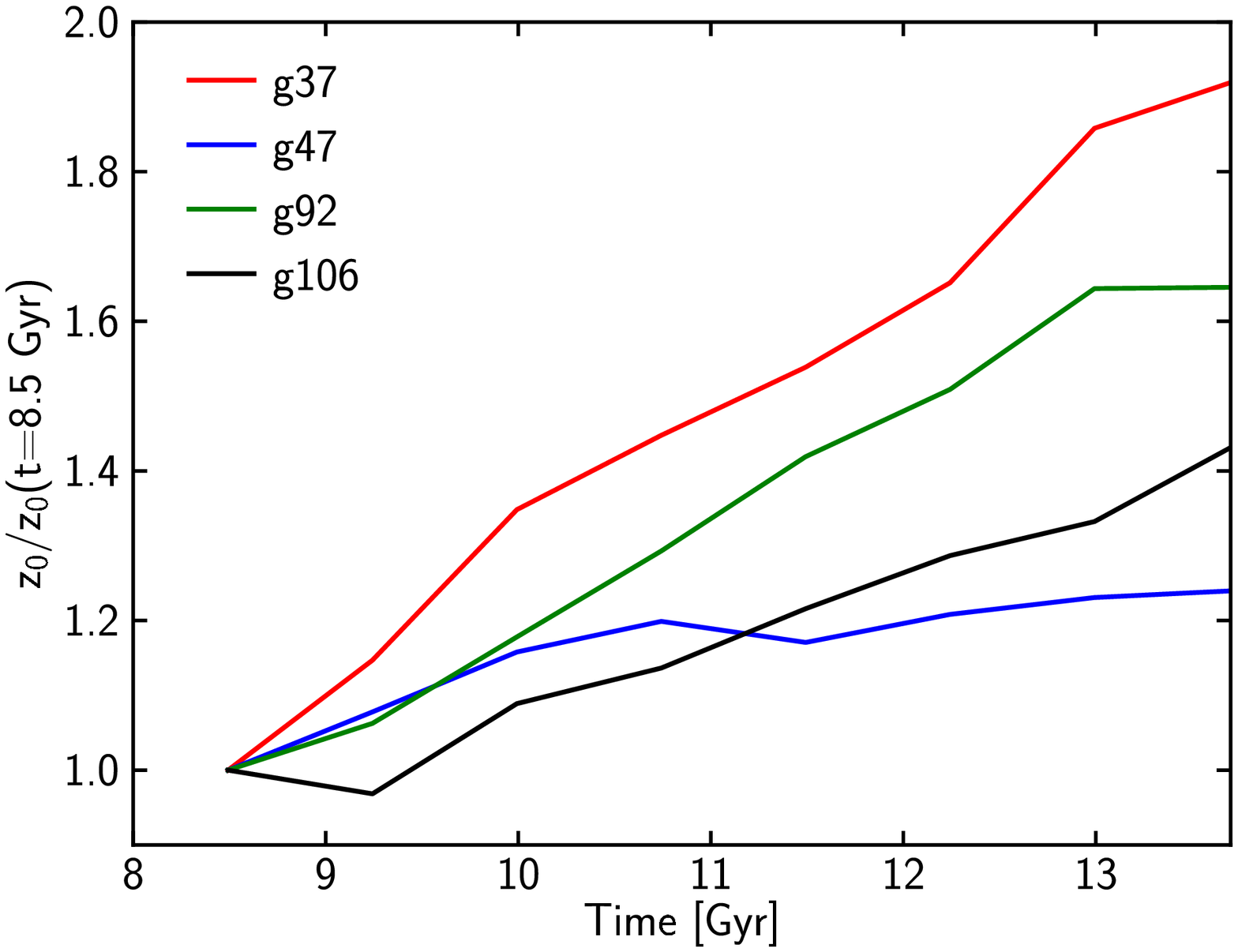}
\includegraphics[width=0.33\textwidth]{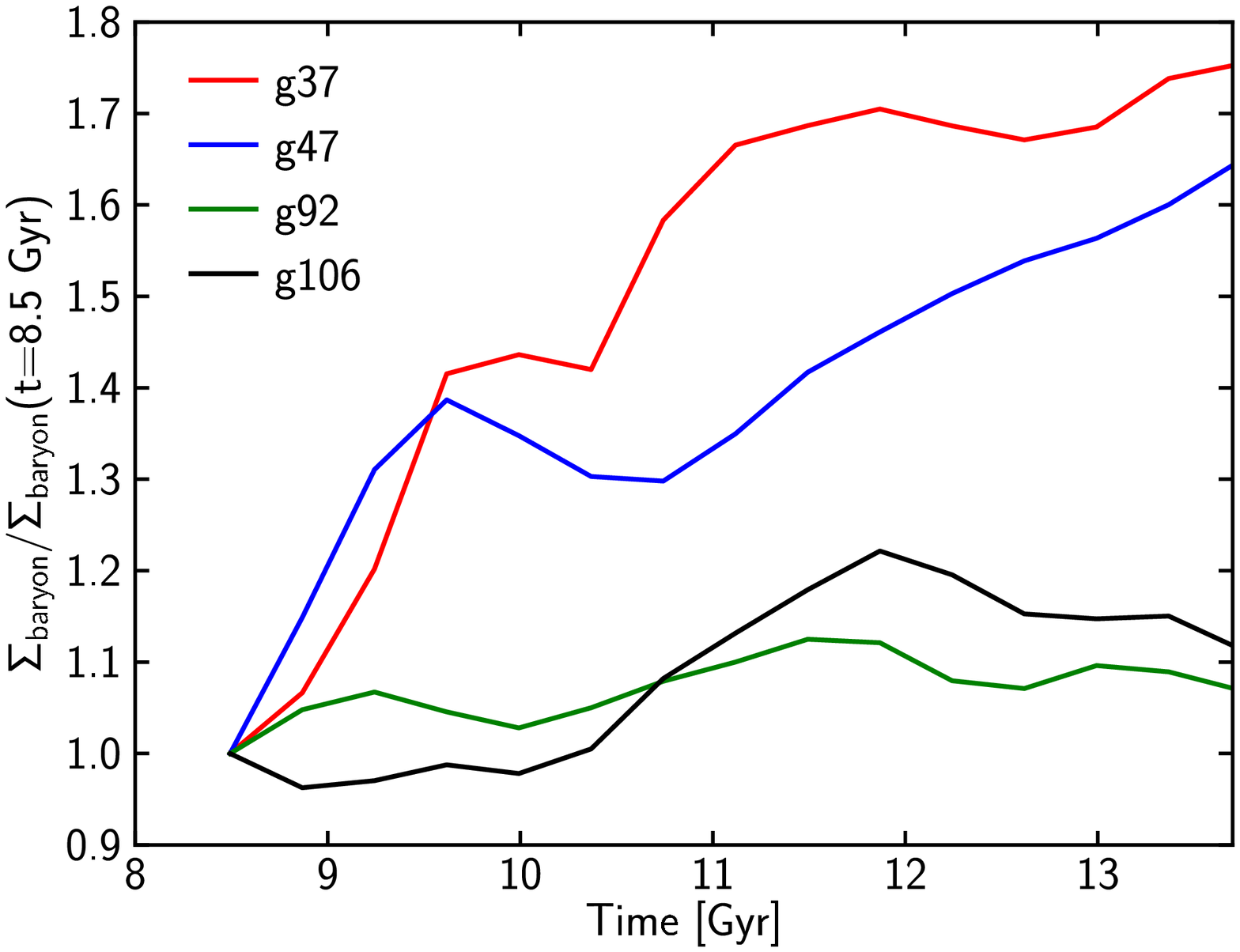}
\caption{Left: time evolution of $\sigma_z^2$ for one mono-age population in each of the four simulated galaxies (the slope of each line is the heating rate $D$ as defined in Section 4). Middle:  evolution of the scaleheight. Right: evolution of the baryonic surface density (gas+stars) at 2$R_d$. In thin dashed lines in the left panel is shown the heating that can be attributed to the increase in surface density, computed using $\sigma_z^2(t)=2 \pi G \Sigma(t) z_0$($t=8.5$ Gyr). We find that disc growth only has a minor contribution to disc heating for g92 and g106, but is much more significant in g37 and g47.}
\label{fig:hydrostatic}
\end{figure*}
To elucidate the mechanisms for heating, for each of the four galaxies of the previous section, we study the time evolution of one mono-age population, composed of the stars born at times between 8 and 8.5 Gyr, and that are found in a ring around 2$R_d$ at $t=13.7$ Gyr.

The first possible mechanism we study is simply heating due to disc growth (i.e., mass accretion). Indeed, from the vertical Jeans equation for an  isothermal sheet with a flat rotation curve, one can see that the vertical velocity dispersion depends on the disc surface density and the scale-height \citep{vanderKruit1988}:
$$
\sigma_z=\sqrt{2 \pi G \Sigma z_0}
$$
Consequently, in a disc with an increasing surface density, $\sigma_z$ is expected to increase with time (under the conservative assumption that the scale-height remains constant). This is an effect that is not traditionally considered in papers studying heating (these papers usually focus on pre-assembled galaxies, whose mass does not evolve), although it has already been shown by \cite{Jenkins1992} and Villalobos et al (2010) that an increase of the mid-plane density leads to adiabatic heating in the disc.

We show in Figure \ref{fig:hydrostatic} the time evolution of the baryonic surface density (i.e., including both gas and stars, which both contribute to the gravity in the disc plane) within an annulus at 2$R_d$ (up to $z$ of 3 kpc). We find that over the 5 Gyr that we consider, the surface density only increases by $\sim$10\% for g106 and g92, but it increases by 65 to 75\% for the other two galaxies. This could be an important source of heating. The exact contribution to the increase of $\sigma_z$ actually depends on how the scale-height of the population evolves with time (middle panel, Figure \ref{fig:hydrostatic}). In all cases, $z_0$ increases with time, with large differences between galaxies.

This increase of $z_0$ with time is unexpected if the vertical action is conserved (which should be the case for adiabatic changes of the gravitational potential). Indeed, using the epicycle approximation, the vertical action can be written as (see e.g., \citealp{Solway2012})
$$
J_{z,epi}=\frac{E_{z,epi}}{\nu},
$$
where $E_{z,epi}$ is an approximation to the vertical energy
$$
E_{z,epi}=\frac{1}{2}v_z^2+\frac{1}{2}\nu^2 z^2
$$
and $\nu$, the vertical epicyclic frequency, is proportional to $\sqrt{\rho_0}$, where $\rho_0$ is the mid-plane density. Consequently, to conserve the vertical action, when the disc density increases and \sz increases, then the scale-height has to decrease. The fact that we measure an increase of $z_0$ then means that the vertical action is not being conserved as a function of time. This could be due either to a non-adiabatic growth of the disc mass or to non-axisymmetric features. Note that \cite{Solway2012} find that the vertical action is not perfectly conserved, even for simulated discs that do not accrete mass with time (despite computing the action very carefully). For a growing disc, \cite{Sridhar1996} have  shown that only the sum of the vertical and radial action is conserved, and not the vertical action itself (see also \citealp{Minchev2012}).

This makes it difficult to estimate the contribution of gravity-driven heating to the increase of \sz (and even more so because it is not the only heating mechanism present as explained in the next Section). We provide  a rough estimate of  how much of the heating can be attributed to the increased surface density, by computing a gravity-driven $\sigma_z$, using the values of the surface density as a function of time, and assuming that the scale-height remains fixed at its initial value. 

These are the dashed lines in Figure \ref{fig:hydrostatic}, which show that the increase in surface density contributes to the heating rate by only 5\% for g92 and 14\% for g106 (their surface density evolves very little). On the other hand, it contributes by 37\% and 57\% to the heating in g37 and g47, respectively.

Since these contributions are sub-dominant, in the next section, we investigate additional sources of heating. 

\subsection{Spiral arms, bar and bending modes}
\begin{figure*}
\centering
\includegraphics[width=0.33\textwidth]{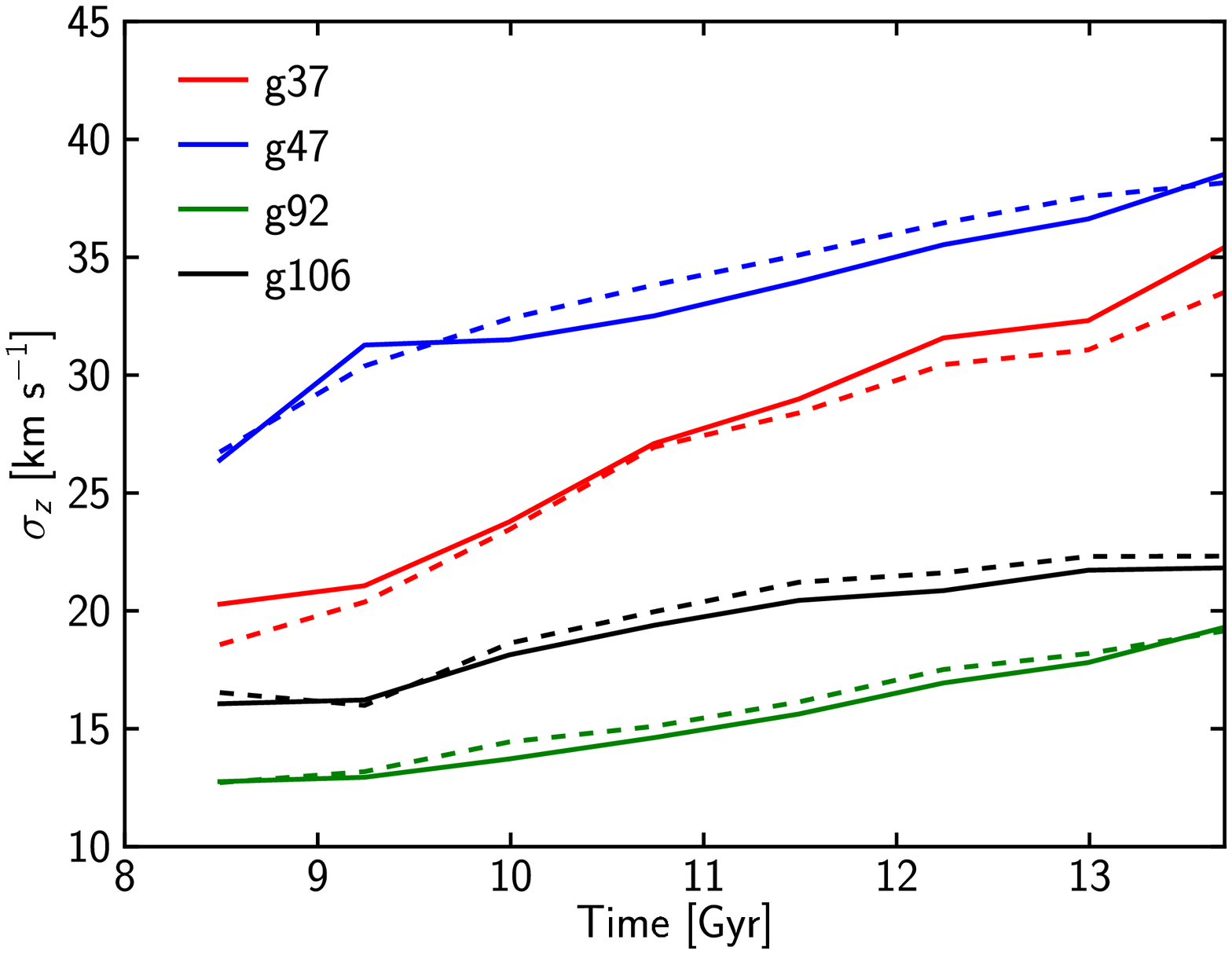}
\includegraphics[width=0.33\textwidth]{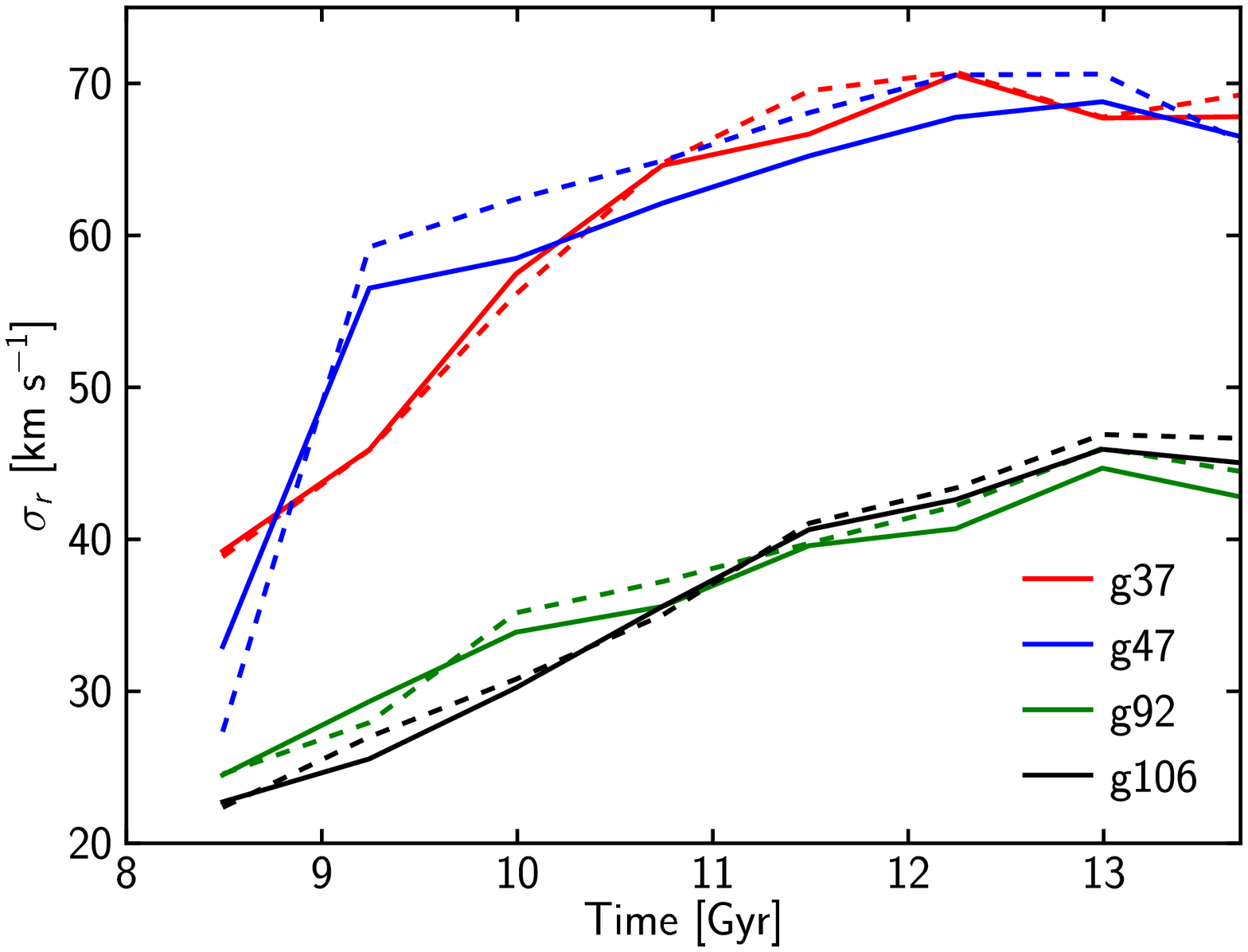}
\includegraphics[width=0.33\textwidth]{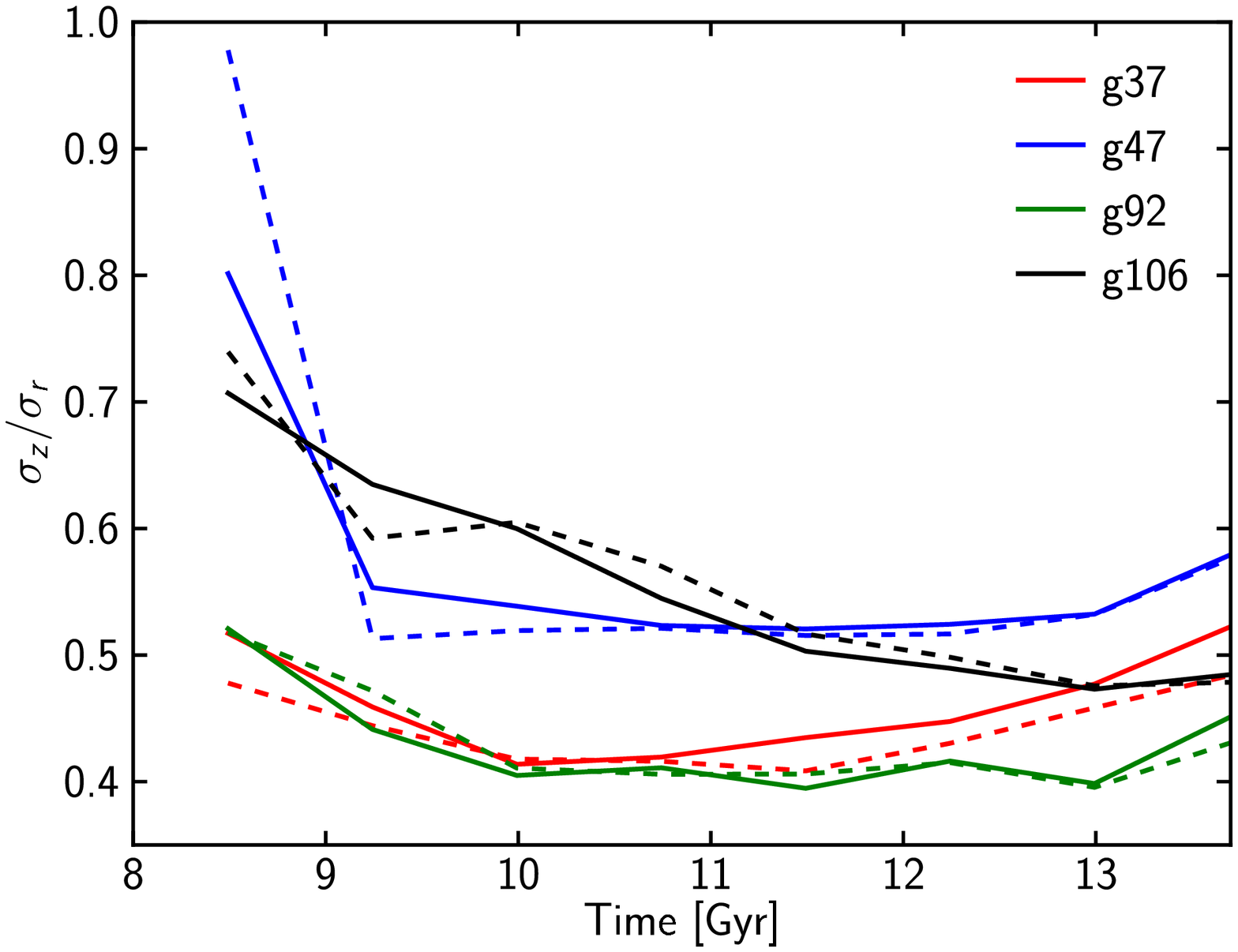}
\caption{Time evolution of the vertical velocity dispersion ($\sigma_z$, left panel), radial velocity dispersion ($\sigma_r$, middle panel), and of the ratio between the two ($\sigma_z/\sigma_r$, right panel), for the 4 galaxies with quiescent histories. We show for each galaxy the velocity dispersion for a population of stars born between 8 and 8.5 Gyr, and contained within an annulus at 2$R_d$ at $t=13.7$ Gyr (up to $z=3$ kpc). The solid lines represent all stars, irrespective of their birth location, while the dashhed line corresponds to stars born within the final 2$R_d$ annulus (and thus corresponds to stars with minimal radial migration).}
\label{fig:sigma_evol}
\end{figure*}

A possible diagnostic of heating mechanisms is the ratio of vertical to radial velocity dispersions. In Figure \ref{fig:sigma_evol} we show the time evolution of \sz, of \sr and of $\sigma_z/\sigma_r$ for the populations  studied in the previous section (stars born between $t=8$ and 8.5 Gyr). In all galaxies, we find a steady increase of $\sigma_z$ with time, with the possible exception of g106 where $\sigma_z$ seems to saturate to $\sim 22$ \kms. We similarly find that $\sigma_r$ increases with time (with a saturation at  $\sim 65$ \kms for g37 over the last 3 Gyr of evolution). The resulting $\sigma_z/\sigma_r$ ratio has a complex behaviour, suggesting different sources of heating from one galaxy to another (and confirming that most of the heating is not simply numerical).

For g47, g92 and g106, $\sigma_z/\sigma_r$  decreases over the 5 Gyr that we consider. For g37, $\sigma_z/\sigma_r$  first decreases for 1.5 Gyr, and then returns to its initial value.  These patterns are not consistent with numerical heating: \cite{Sellwood2013} shows that $\sigma_z/\sigma_r$  increases strongly for disc undergoing numerical heating (and in simulations with large number of particles, $\sigma_z/\sigma_r$ evolves very little). A decrease of $\sigma_z/\sigma_r$ is also inconsistent with the models of heating by GMCs and halo black holes by \cite{Hanninen2002}, where $\sigma_z/\sigma_r$ is either constant or slightly increasing with time.

Spiral arms and bars are more efficient at heating discs in the radial direction \citep{Sellwood1984, Carlberg1985,  Minchev2006}. Vertical heating can then be obtained if the radial heating is redistributed in the vertical direction by GMCs or other massive objects \citep{Carlberg1987, Jenkins1990, Jenkins1992}. \cite{Saha2010} also observe vertical heating linked with the presence of spiral arms, and speculate that even in the absence of GMCs the radial heating could be redistributed in the vertical direction by weak bending waves (as proposed by \citealp{Masset1997}).

\begin{figure}
\centering
\includegraphics[width=0.45\textwidth]{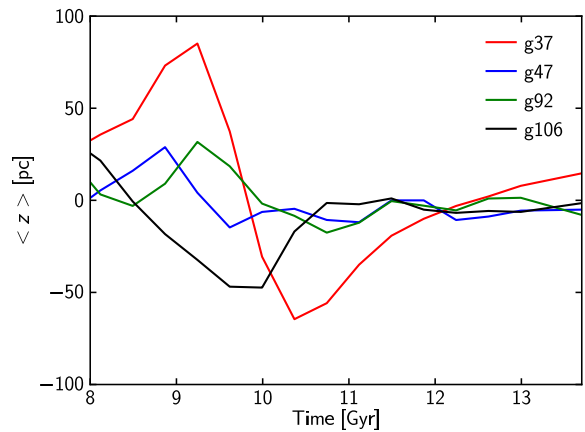}
\caption{Time evolution on the average height above the plane for all stellar particles at  2$R_d$ (up to $z$ of 1 kpc). All galaxies show some oscillations (corresponding to bending modes), which are strongest for g37.}
\label{fig:zmean}
\end{figure}

A very likely source of heating in our simulations could then be the combination of spiral arms with either overdensities in the disc and/or weak bending waves. We show in Figure \ref{fig:zmean} that such bending waves are indeed present in the simulations. We show the mean height above the mid-plane of stellar particles as a function of time  for an annulus at 2$R_d$ in each simulated disc, and we find low levels of vertical oscillations for all galaxies.

Interestingly, these bending waves are much stronger for g37, for which the evolution of $\sigma_z/\sigma_r$ is also different. In this case, we suspect that the late increase of $\sigma_z/\sigma_r$ could be due to the bending waves themselves \citep{Khoperskov2010,Griv2011}. \cite{Khoperskov2010} also show that bending waves increase $\sigma_z/\sigma_r$ mostly through an increase of \sz, which is what we find for g37 where at late times \sz increases while \sr remains approximately constant (note that this was also observed by \citealp{Saha2010}). The bending waves themselves could be due to small satellites orbiting the simulated galaxies \citep{Widrow2014}, but could also naturally arise from internal instabilities (that are predicted to occur when $\sigma_z/\sigma_r$ is below $\sim$ 0.4), or even be induced by the spiral arms \citep{Faure2014}.

\vspace{1cm}
We conclude from the analysis in this section that disc heating in our simulations is  due neither to numerical effects, nor to radial migration. The contributions to the heating vary from one galaxy to another, and include heating due to disc growth, and heating due to a combination of spiral arms and  bars coupled with overdensities in the disc and vertical bending waves.

\section{Summary}
We study the age-velocity relation in a sample of seven simulated disc galaxies. All galaxies first undergo an active phase of mergers at high redshift, creating in all cases a thick stellar component with a high velocity dispersion. Amongst the old stars, we find that most of them are actually born kinematically hot, with an additional contribution of accreted stars.
At lower redshift (corresponding to stellar ages smaller than 8--9 Gyr), three of the simulated galaxies have a quiescent history while the others have various types of mergers. We find that in the quiescent galaxies  \sz increases smoothly with age. When fitted with a power law, we find heating indices close to 0.5, making our simulated galaxies consistent with results from the Geneva-Copenhagen Survey for the solar neighbourhood \citep{Nordstrom2004,Holmberg2007}, rather than with observations suggesting that disc  heating saturates after a few Gyrs \citep[e.g.,][]{Soubiran2008}. Note that such a gradual heating was also found in other zoom cosmological simulations, as for instance shown by \cite{Brook2012} or \cite{Bird2013}, even if they do not measure the heating index in their simulations.

By contrast with \cite{Bird2013}, though, we mostly find that the slope of the AVR is not imprinted at birth, but is due to heating. While we cannot pin down the exact mechanisms responsible for that heating, we show that radial migration is not one of them. Because of the combined contribution of hotter stars coming from the inner disc and cooler stars coming from the outer disc, we even find that the total effect of migration is to very slightly cool the disc, at least for the radius we examined, and for stars younger than $\sim$ 8 Gyr. For older stars, the effect is reversed, because stars coming from the inner disc are cooler: this is due to early mergers creating a flared structure for the old stars.

Finally, we find that \sz is very sensitive to mergers \citep[see also e.g., ][]{Quinn1993,Velazquez1999,Kazantzidis2009, House2011}: even 1--10 to 1--15 mergers create jumps in the age-velocity relation. This can in theory be used to probe the merger history of the Milky Way. However, we find that errors of 30\% on stellar ages, even if they do not affect the shape of the AVR for quiescent galaxies (except for the oldest stars, for which age errors significantly decrease the measured \sz), can significantly blur the signatures of mergers (errors of 20\% or less are necessary to properly detect these signatures). Stronger constraints on the history of the Milky Way can be obtained by combining the AVR with the structure of mono-abundance populations \citep{Bovy2012b,Bovy2012c}. As discussed in Paper I, our simulations, combined with current data, favour a very quiescent history for the Milky Way in the last 9 Gyr, with a potential contribution of mergers to the build-up of the thickest and oldest disc components.

\section*{Acknowledgments}
We thank the referee for a constructive report.
We thank Hans-Walter Rix and Jo Bovy for useful comments. MM is supported by a Humboldt Research Fellowship.
CF acknowledges financial support by the Beckwith Trust.

{}


\begin{thebibliography}{plain}

\bibitem[\protect\citeauthoryear{Agertz et al.}{2009}]{Agertz2009} Agertz O., Lake G., Teyssier R., Moore B., Mayer L., Romeo A.~B., 2009, MNRAS, 392, 294 
\bibitem[\protect\citeauthoryear{Agertz et al.}{2013}]{Agertz2013}Agertz O., Kravtsov A.~V., Leitner S.~N., Gnedin N.~Y., 2013, ApJ, 770, 25 
\bibitem[\protect\citeauthoryear{Aumer \& Binney}{2009}]{Aumer2009} Aumer M., Binney J.~J., 2009, MNRAS, 397, 1286 
\bibitem[\protect\citeauthoryear{Bird et al.}{2013}]{Bird2013} Bird J.~C., Kazantzidis S., Weinberg D.~H., Guedes J., Callegari S., Mayer L., Madau P., 2013, ApJ, 773, 43 
\bibitem[\protect\citeauthoryear{Bournaud, Elmegreen, \& Martig}{2009}]{Bournaud2009} Bournaud F., Elmegreen B.~G., Martig M., 2009, ApJ, 707, L1 
\bibitem[\protect\citeauthoryear{Bournaud et al.}{2010}]{Bournaud2010} Bournaud F., Elmegreen B.~G., Teyssier R., Block D.~L., Puerari I., 2010, MNRAS, 409, 1088 
\bibitem[Bovy et al.(2012a)]{Bovy2012a} Bovy J., Rix H.-W., Hogg D.~W.\ 2012a, ApJ, 751, 131 
\bibitem[Bovy et al.(2012b)]{Bovy2012b} Bovy J., Rix H.-W., Liu C., Hogg D. W., Beers T. C., Lee, Y. S.\ 2012b, ApJ, 753, 148 
\bibitem[Bovy et al.(2012c)]{Bovy2012c} Bovy J., Rix H.-W., Hogg D.~W., Beers T. C., Lee, Y. S., Zhang L. \ 2012c, ApJ, 755, 115 
\bibitem[\protect\citeauthoryear{Brook et al.}{2012}]{Brook2012} Brook C.~B., et al., 2012, MNRAS, 426, 690 
\bibitem[\protect\citeauthoryear{Carlberg \& Sellwood}{1985}]{Carlberg1985} Carlberg R.~G., Sellwood J.~A., 1985, ApJ, 292, 79 
\bibitem[\protect\citeauthoryear{Carlberg}{1987}]{Carlberg1987} Carlberg R.~G., 1987, ApJ, 322, 59 
\bibitem[\protect\citeauthoryear{Casagrande et al.}{2011}]{Casagrande2011} Casagrande L., Sch{\"o}nrich R., Asplund M., Cassisi S., Ram{\'{\i}}rez I., Mel{\'e}ndez J., Bensby T., Feltzing S., 2011, A\&A, 530, A138 
\bibitem[\protect\citeauthoryear{Dib, Bell, \& Burkert}{2006}]{Dib2006} Dib S., Bell E., Burkert A., 2006, ApJ, 638, 797 
\bibitem[Faure et al.(2014)]{Faure2014} Faure, C., Siebert, A., Famaey, B.\ 2014, arXiv:1403.0587 
\bibitem[\protect\citeauthoryear{Forbes, Krumholz, \& Burkert}{2012}]{Forbes2012} Forbes J., Krumholz M., Burkert A., 2012, ApJ, 754, 48 
\bibitem[\protect\citeauthoryear{F{\"o}rster Schreiber et al.}{2009}]{Forster2009} F{\"o}rster Schreiber N.~M., et al., 2009, ApJ, 706, 1364 
\bibitem[\protect\citeauthoryear{Governato et al.}{2004}]{Governato2004} Governato F., et al., 2004, ApJ, 607, 688 
\bibitem[Griv (2011)]{Griv2011} Griv, E.\ 2011, MNRAS, 415, 1259 
\bibitem[\protect\citeauthoryear{H{\"a}nninen \& Flynn}{2002}]{Hanninen2002} H{\"a}nninen J., Flynn C., 2002, MNRAS, 337, 731 
\bibitem[\protect\citeauthoryear{Haywood et al.}{2013}]{Haywood2013} Haywood M., Di Matteo P., Lehnert M., Katz D., Gomez A., 2013, arXiv, arXiv:1305.4663 
\bibitem[\protect\citeauthoryear{Holmberg, Nordstr{\"o}m, \& Andersen}{2007}]{Holmberg2007} Holmberg J., Nordstr{\"o}m B., Andersen J., 2007, A\&A, 475, 519 
\bibitem[House et al.(2011)]{House2011} House E.~L. et al.\, 2011, MNRAS, 415, 2652 
\bibitem[Inoue \& Saitoh(2014)]{Inoue2014} Inoue, S.,  Saitoh, T.~R.\ 2014, arXiv:1402.5986 
\bibitem[\protect\citeauthoryear{Jenkins \& Binney}{1990}]{Jenkins1990} Jenkins A., Binney J., 1990, MNRAS, 245, 305 
\bibitem[Jenkins (1992)]{Jenkins1992} Jenkins, A.\ 1992, MNRAS, 257, 620 
\bibitem[\protect\citeauthoryear{Kaufmann et al.}{2007}]{Kaufmann2007} Kaufmann T., Mayer L., Wadsley J., Stadel J., Moore B., 2007, MNRAS, 375, 53 
\bibitem[\protect\citeauthoryear{Kazantzidis et al.}{2009}]{Kazantzidis2009} Kazantzidis S., Zentner A.~R., Kravtsov A.~V., Bullock J.~S., Debattista V.~P., 2009, ApJ, 700, 1896 
\bibitem[\protect\citeauthoryear{Khoperskov et al.}{2007}]{Khoperskov2007} Khoperskov A.~V., Just A., Korchagin V.~I., Jalali M.~A., 2007, A\&A, 473, 31 
\bibitem[Khoperskov et al. (2010)]{Khoperskov2010} Khoperskov, A., Bizyaev, D., Tiurina, N., Butenko, M.\ 2010, AN, 331, 731 
\bibitem[\protect\citeauthoryear{Lacey}{1984}]{Lacey1984} Lacey C.~G., 1984, MNRAS, 208, 687 
\bibitem[\protect\citeauthoryear{Lacey \& Ostriker}{1985}]{Lacey1985} Lacey C.~G., Ostriker J.~P., 1985, ApJ, 299, 633 
\bibitem[\protect\citeauthoryear{Lee et al.}{2011}]{Lee2011} Lee Y.~S., et al., 2011, ApJ, 738, 187 
\bibitem[\protect\citeauthoryear{Loebman et al.}{2011}]{Loebman2011} Loebman S.~R., Ro{\v s}kar R., Debattista V.~P., Ivezi{\'c} {\v Z}., Quinn T.~R., Wadsley J., 2011, ApJ, 737, 8 
\bibitem[Martig et al. (2009)]{Martig2009} Martig M., Bournaud F., Teyssier R., Dekel A.,\ 2009, ApJ, 707, 250 
\bibitem[Martig \& Bournaud (2010)]{Martig2010} Martig M., Bournaud F.,\ 2010, ApJ, 714, L275 
\bibitem[Martig et al. (2012)]{Martig2012} Martig M., Bournaud F., Croton D.~J., Dekel A., Teyssier R.,\ 2012, ApJ, 756, 26
\bibitem[Martig et al. (2014a)]{Martig2014a} Martig M., Minchev I., Flynn C.,\ 2014, MNRAS,  442, 2474 
\bibitem[Masset \& Tagger (1997)]{Masset1997} Masset, F., \& Tagger, M.\ 1997, A\&A, 322, 442 
\bibitem[\protect\citeauthoryear{Mayer}{2004}]{Mayer2004} Mayer L., 2004, arXiv:astro-ph/0411476  
\bibitem[\protect\citeauthoryear{Minchev \& Quillen}{2006}]{Minchev2006} Minchev I., Quillen A.~C., 2006, MNRAS, 368, 623
\bibitem[\protect\citeauthoryear{Minchev \& Famaey}{2010}]{Minchev2010} Minchev I., Famaey B., 2010, ApJ, 722, 112 
\bibitem[Minchev et al. (2012)]{Minchev2012} Minchev I., Famaey B., Quillen A.~C., Dehnen W., Martig M., Siebert A.,\ 2012, A\&A, 548, 127 
\bibitem[\protect\citeauthoryear{Minchev, Chiappini, \& Martig}{2013}]{Minchev2013} Minchev, I., Chiappini, C., Martig, M.\ 2013, A\&A, 558, 9
\bibitem[Minchev et al. (2014)]{Minchev2014} Minchev, I., et al.\ 2014, ApJ, 781, L20 
\bibitem[\protect\citeauthoryear{Moster et al.}{2010}]{Moster2010} Moster B.~P., Macci{\`o} A.~V., Somerville R.~S., Johansson P.~H., Naab T., 2010, MNRAS, 403, 1009  
\bibitem[\protect\citeauthoryear{Nordstr{\"o}m et al.}{2004}]{Nordstrom2004} Nordstr{\"o}m B., et al., 2004, A\&A, 418, 989 
\bibitem[\protect\citeauthoryear{Pilkington et al.}{2011}]{Pilkington2011} Pilkington K., et al., 2011, MNRAS, 417, 2891
\bibitem[\protect\citeauthoryear{Quillen \& Garnett}{2001}]{Quillen2001} Quillen A.~C., Garnett D.~R., 2001, ASPC, 230, 87  
\bibitem[\protect\citeauthoryear{Quinn, Hernquist, \& Fullagar}{1993}]{Quinn1993} Quinn P.~J., Hernquist L., Fullagar D.~P., 1993, ApJ, 403, 74 
\bibitem[\protect\citeauthoryear{Renaud et al.}{2013}]{Renaud2013} Renaud F., et al., 2013, arXiv, arXiv:1307.5639 
\bibitem[\protect\citeauthoryear{Ro{\v s}kar, Debattista, \& Loebman}{2013}]{Roskar2013} Ro{\v s}kar R., Debattista V.~P., Loebman S.~R., 2013, MNRAS, 1468
\bibitem[\protect\citeauthoryear{Saha, Tseng, \& Taam}{2010}]{Saha2010} Saha K., Tseng Y.-H., Taam R.~E., 2010, ApJ, 721, 1878  
\bibitem[\protect\citeauthoryear{Sch{\"o}nrich \& Binney}{2009a}]{Schonrich2009a} Sch{\"o}nrich R., Binney J., 2009a, MNRAS, 396, 203 
\bibitem[\protect\citeauthoryear{Sch{\"o}nrich \& Binney}{2009b}]{Schonrich2009b} Sch{\"o}nrich R., Binney J., 2009b, MNRAS, 399, 1145 
\bibitem[\protect\citeauthoryear{Seabroke \& Gilmore}{2007}]{Seabroke2007} Seabroke G.~M., Gilmore G., 2007, MNRAS, 380, 1348 
\bibitem[Sellwood \& Carlberg (1984)]{Sellwood1984} Sellwood, J.~A., Carlberg, R.~G.\ 1984, ApJ, 282, 61 
\bibitem[\protect\citeauthoryear{Sellwood \& Binney}{2002}]{Sellwood2002} Sellwood J.~A., Binney J.~J., 2002, MNRAS, 336, 785 
\bibitem[\protect\citeauthoryear{Sellwood}{2013}]{Sellwood2013} Sellwood J.~A., 2013, ApJ, 769, L24 
\bibitem[\protect\citeauthoryear{Soderblom}{2010}]{Soderblom2010} Soderblom D.~R., 2010, ARA\&A, 48, 581 
\bibitem[\protect\citeauthoryear{Solway, Sellwood, \& Sch{\"o}nrich}{2012}]{Solway2012} Solway M., Sellwood J.~A., Sch{\"o}nrich R., 2012, MNRAS, 422, 1363 
\bibitem[\protect\citeauthoryear{Soubiran et al.}{2008}]{Soubiran2008} Soubiran C., Bienaym{\'e} O., Mishenina T.~V., Kovtyukh V.~V., 2008, A\&A, 480, 91 
\bibitem[\protect\citeauthoryear{Spitzer \& Schwarzschild}{1951}]{Spitzer1951} Spitzer L., Jr., Schwarzschild M., 1951, ApJ, 114, 385 
\bibitem[\protect\citeauthoryear{Spitzer \& Schwarzschild}{1953}]{Spitzer1953} Spitzer L., Jr., Schwarzschild M., 1953, ApJ, 118, 106 
\bibitem[Sridhar \& Touma (1996)]{Sridhar1996} Sridhar S.,  Touma J.,\ 1996, Science, 271, 973 
\bibitem[\protect\citeauthoryear{Stark \& Brand}{1989}]{Stark1989} Stark A.~A., Brand J., 1989, ApJ, 339, 763 
\bibitem[\protect\citeauthoryear{Stark \& Lee}{2005}]{Stark2005} Stark A.~A., Lee Y., 2005, ApJ, 619, L159 
\bibitem[\protect\citeauthoryear{Stark \& Lee}{2006}]{Stark2006} Stark A.~A., Lee Y., 2006, ApJ, 641, L113 
\bibitem[\protect\citeauthoryear{Steinmetz \& White}{1997}]{Steinmetz1997} Steinmetz M., White S.~D.~M., 1997, MNRAS, 288, 545 
\bibitem[\protect\citeauthoryear{Tamburro et al.}{2009}]{Tamburro2009} Tamburro D., Rix H.-W., Leroy A.~K., Mac Low M.-M., Walter F., Kennicutt R.~C., Brinks E., de Blok W.~J.~G., 2009, AJ, 137, 4424 
\bibitem[\protect\citeauthoryear{Vande Putte, Cropper, \& Ferreras}{2009}]{VandePutte2009} Vande Putte D., Cropper M., Ferreras I., 2009, MNRAS, 397, 1587 
\bibitem[van der Kruit (1988)]{vanderKruit1988} van der Kruit, P.~C.\ 1988, A\&A, 192, 117 
\bibitem[van Dokkum et al. (2013)]{vanDokkum2013} van Dokkum, P.~G.,  et al.\ 2013, ApJ, 771, L35 
\bibitem[\protect\citeauthoryear{Velazquez \& White}{1999}]{Velazquez1999} Velazquez H., White S.~D.~M., 1999, MNRAS, 304, 254 
\bibitem[\protect\citeauthoryear{Villalobos \& Helmi}{2008}]{Villalobos2008} Villalobos {\'A}., Helmi A., 2008, MNRAS, 391, 1806 
\bibitem[\protect\citeauthoryear{Wada, Meurer, \& Norman}{2002}]{Wada2002} Wada K., Meurer G., Norman C.~A., 2002, ApJ, 577, 197 
\bibitem[\protect\citeauthoryear{Walker, Mihos, \& Hernquist}{1996}]{Walker1996} Walker I.~R., Mihos J.~C., Hernquist L., 1996, ApJ, 460, 121 
\bibitem[\protect\citeauthoryear{Widrow \& Dubinski}{2005}]{Widrow2005} Widrow L.~M., Dubinski J., 2005, ApJ, 631, 838 
\bibitem[Widrow et al. (2014)]{Widrow2014} Widrow L.~M., Barber J., Chequers M.~H.,  Cheng  E.\ 2014, MNRAS, 440, 1971 
\bibitem[\protect\citeauthoryear{Wielen}{1977}]{Wielen1977} Wielen R., 1977, A\&A, 60, 263  
\bibitem[\protect\citeauthoryear{Wilson et al.}{2011}]{Wilson2011} Wilson C.~D., et al., 2011, MNRAS, 410, 1409 
\end{thebibliography}
\end{document}